\documentclass[twocolumn,twocolappendix]{aastex701}

\shorttitle{Eps Ind Ab ammonia and more}
\shortauthors{Matthews et al.}
\accepted{March 8, 2026}

\usepackage{booktabs}
\newcommand{\Mjup}{\ensuremath{M_{\text{Jup}}}}
\usepackage{xcolor}
\usepackage[version=4]{mhchem}

\begin{document}

\title{A second visit to Eps~Ind~Ab with JWST: new photometry confirms ammonia and suggests thick clouds in the exoplanet atmosphere of the closest super-Jupiter.}

\author[0000-0003-0593-1560]{Elisabeth C. Matthews} 
\affiliation{Max Planck Institute for Astronomy, K{\"o}nigstuhl 17, D-69117 Heidelberg, Germany}
\email[show]{matthews@mpia.de} 
\correspondingauthor{Elisabeth C. Matthews}

\author[0000-0001-5864-9599]{James Mang} 
\altaffiliation{NSF Graduate Research Fellow.}
\affiliation{Department of Astronomy, University of Texas at Austin, 2515 Speedway, Austin, TX 78712, USA}
\email{j_mang@utexas.edu} 

\author[0000-0001-5365-4815]{Aarynn L. Carter} 
\affiliation{Space Telescope Science Institute, 3700 San Martin Drive, Baltimore, MD 21218, USA}
\email{aacarter@stsci.edu} 

\author[0000-0002-2918-8479]{Mathlide M\^alin} 
\affiliation{Space Telescope Science Institute, 3700 San Martin Drive, Baltimore, MD 21218, USA}
\email{mmalin@stsci.edu} 

\author[0000-0002-4404-0456]{Caroline V. Morley} 
\affiliation{Department of Astronomy, University of Texas at Austin, 2515 Speedway, Austin, TX 78712, USA}
\email{cmorley@utexas.edu} 

\author[0009-0004-9729-6377]{Bhavesh Rajpoot} 
\affiliation{Max Planck Institute for Astronomy, K{\"o}nigstuhl 17, D-69117 Heidelberg, Germany}
\affiliation{Department of Physics and Astronomy, Heidelberg University, Im Neuenheimer Feld 226, D-69120 Heidelberg, Germany}
\email{rajpoot@mpia.de} 


\author[0000-0002-3952-8588]{Leindert A. Boogaard} 
\affiliation{Leiden Observatory, Leiden University, PO Box 9513, NL-2300 RA Leiden, The Netherlands}
\email{boogaard@strw.leidenuniv.nl}

\author[0000-0002-0040-6815]{Jennifer A. Burt}
\affiliation{Jet Propulsion Laboratory, California Institute of Technology, 4800 Oak Grove Drive, Pasadena, CA 91109, USA}
\email{jennifer.burt@jpl.nasa.gov} 

\author[0000-0000-0000-0000]{Ian J. M. Crossfield}
\affiliation{Department of Physics and Astronomy, University of Kansas, Lawrence, KS, USA}
\email{ianc@ku.edu} 

\author[0000-0001-6039-0555]{Fabo Feng}
\affiliation{Shanghai Jiao Tong University, Shengrong Road 520, Shanghai, 201210, People’s Republic Of
China}
\email{ffeng@sjtu.edu.cn} 

\author[0000-0002-2189-2365]{Anne-Marie Lagrange}
\affiliation{LESIA, Observatoire de Paris, Université PSL, CNRS, 5 Place Jules Janssen, 92190 Meudon, France}
\email{Anne-marie.Lagrange@obspm.fr}

\author[0000-0001-6041-7092]{Mark W Phillips}
\affiliation{Institute for Astronomy, University of Edinburgh, Royal Observatory, Blackford Hill, Edinburgh EH9 3HJ, UK}
\email{mark.phillips@roe.ac.uk}

\begin{abstract}
With JWST, we are directly imaging cold ($\sim$200-300K), solar-age giant exoplanets for the first time. At these temperatures many molecular features appear and water-ice clouds may condense and affect the emission spectrum; early photometric measurements of cold giant planets are already showing some tension with the predictions of cloud-free, solar-metallicity atmosphere models.
Here we present new JWST/MIRI coronagraphic observations of the cold giant exoplanet Eps~Ind~Ab at 11.3\micron. Together with archival data, we use these new observations to study the atmosphere of this cold exoplanet, and we also re-fit its orbit, finding an updated mass of $7.6\pm0.7$\Mjup~and an eccentricity of $0.24^{+0.11}_{-0.08}$. 
The planet is significantly brighter (by $0.88\pm0.08$\,mag) at 11.3\micron~than at 10.6\micron, indicating the presence of ammonia. However, this ammonia feature is shallower than expected. This could indicate a low-metallicity or nitrogen-depleted atmosphere, but our preferred explanation is the presence of thick water-ice clouds that suppress the ammonia feature and the near-IR emission of Eps Ind Ab. 
Photometry of the small but growing sample of cold, giant exoplanets demonstrates that they are consistently fainter than expected between 3 to 5\micron, consistent with the water-ice cloud hypothesis. 10.6\micron~and 11.3\micron~photometry of this cold exoplanet sample would be valuable to determine whether the suppressed ammonia feature is universal, and to frame a new open question about the underlying physical cause.
\end{abstract}


\section{Introduction} 

Through high-contrast imaging of nearby stars, it is possible to study the atmospheric abundances of their long-period exoplanets with photometric and spectroscopic campaigns. The James Webb Space Telescope (JWST), with its suite of coronagraphic instruments \citep{Green2005,Krist2009,Perrin2018,Rouan2000,Boccaletti2022}, provides a number of advantages in this regard. JWST allows for exquisite photometric precision, and it provides access to the mid-infrared -- including the 10-11\micron~region where ammonia has significant opacity. Further, JWST uniquely opens the possibility to image cold exoplanets ($<$500\,K), which are not currently accessible with ground-based imaging \citep[e.g.][]{Bowler2016}. To date, three of these cold exoplanets around main-sequence stars have been imaged with JWST: Eps Ind~Ab \citep{Matthews2024}, 14~Her~c \citep{BardalezGagliuffi2025}, and the candidate planet TWA~7b \citep{Lagrange2025,Crotts2025}. A few free-floating brown dwarfs \citep[e.g][]{Cushing2011,Luhman2014} have also been imaged in this temperature regime.

At these cold temperatures, strong ammonia features are expected, as has already been observed spectroscopically in cold brown dwarfs with Spitzer and JWST \citep[e.g.][]{Roellig2004,Suarez2022,Beiler2024b}. Considering exoplanet imaging, the JWST/MIRI coronagraphs are designed such that the narrowband F1065C and F1140C filters (wavelengths 10.6\micron~and 11.3\micron) sit inside and outside the prominent 10-11\micron~ammonia absorption feature, respectively. Photometry in both filters is thus sufficient to confirm the presence of ammonia in a cold exoplanet, and measure its abundance. This has been demonstrated on-sky in observations of GJ~504~b \citep{Malin2025}, which revealed the planet to have F1065C-F1140C\,$=0.45\pm0.14$\,mag. \citet{Malin2025} used these infrared measurements and archival data to confirm the first confident ($>5\sigma$) detection of ammonia in an exoplanet. The ammonia feature is deeper in the mid-IR than the near-IR, and therefore appears already at warmer temperatures \citep[see~e.g.][]{Suarez2022}, but attempts have also been made to detect ammonia with near-IR observations \citep{Whiteford2023,Welbanks2024}.

The focus of the current work is Eps~Ind~Ab, a $\sim$275\,K planet first imaged by \citet{Matthews2024}. That work presented exoplanet photometry at 10.6\micron~and 15.5\micron~(from JWST), as well as a low-significance photometric detection in a broadband 10.5-12.5\micron~VISIR/NEAR filter. Curiously, the exoplanet was not detected in archival VLT/NaCo observations \citep{Janson2009,Viswanath2021}, and this faint near-IR (3-5\micron) flux is in tension with the predictions of most atmosphere models \citep{Phillips2020,Marley2021}. Sonora ElfOwl models \citep{Mukherjee2024} with high metallicity and high C/O were found to be consistent with all in-hand photometry and upper limits of \citet{Matthews2024}, but this composition would be somewhat surprising for a planet as massive as Eps~Ind~Ab since it requires a significant heavy-element content in the disk and efficient accretion of those heavy elements onto the planet. Planet formation studies predict C/O ratios $\lesssim1.5\times$ solar and metallicities within a few times solar for giant planets \citep[e.g][]{Oberg2011,Molliere2022,Penzlin2024} and previous works have also found that more massive planets have metallicities closer to those of their host stars \citep[e.g.][]{Thorngren2016}.

In this paper, we present a new epoch of JWST/MIRI coronagraphic observations of Eps Ind A. These were collected $\sim$2 years after the first JWST observations, and in a new photometric filter (F1140C). We describe the new observations and data reduction in Sections \ref{sec:observations} and \ref{sec:datareduction}. Combining the new and archival photometry, we see clear evidence for ammonia in the Eps~Ind~Ab atmosphere (Section \ref{sec:ammonia}). The ammonia feature is smaller than expected based on cloud-free, solar-metallicity models, and that the cold brown dwarf WISE 0855 has a similarly small ammonia feature: possibly indicating a universal trend in how ammonia features appear in the coldest substellar atmospheres and posing an open question as to the underlying physics causing this shallow ammonia feature; we explore several hypotheses in Section \ref{sec:ammonia}. We also demonstrate that the companion shows common proper motion with the host star (Section \ref{sec:cpm}), as already suggested in \citet{Matthews2024}, and re-fit the orbit of Eps~Ind~Ab (Section \ref{sec:orbits}). Finally, in Section \ref{sec:discussion} we provide some discussion and concluding thoughts.

\section{Observations}
\label{sec:observations}

We observed Eps Ind A with the JWST/MIRI coronagraphic imager \citep{Boccaletti2015,Boccaletti2022} on 2025-05-10 (Program GO \#5037, PI Matthews), using the F1140C filter and the four-quadrant phase-mask coronagraph (4QPM; \citealt{Rouan2000}). We collected observations of (a) the science target, (b) a PSF reference target so as to subtract the stellar PSF from observations with reference differential imaging (RDI; \citealt{Smith1984}), using the 5-point small grid dither technique \citep{Soummer2013,Lajoie2016}, and (c) two background fields so as to subtract the ``glow stick" stray-light feature \citep{Boccaletti2022} that appears in MIRI coronagraphic images. Observations of the two background field locations were designed to match the integration times of the science and PSF reference observations, and the fields are close to the science and PSF target respectively. Our chosen PSF target was DI Tuc (HD211055): this target was also used as a PSF reference star for the Eps Ind A observations collected in \citet{Matthews2024} and was demonstrated to be a suitable PSF reference target in that work. Full details of the observation sequence are provided in Table \ref{tab:jwstobservations}, and all JWST observations of Eps Ind Ab used in this paper can be found in MAST at \dataset[https://doi.org/10.17909/zep8-1p53]{https://doi.org/10.17909/zep8-1p53}.

\begin{figure}
    \centering
    \includegraphics[width=0.5\textwidth]{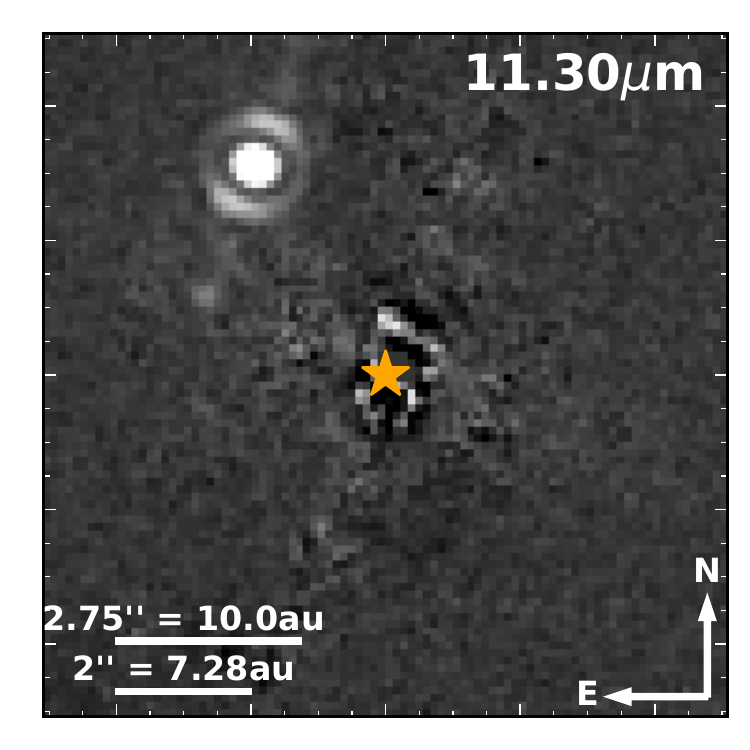}
    \caption{Coronagraphic images of Eps Ind A, collected with the F1140C filter of JWST/MIRI. The planet is detected as a bright point source in upper left of this image.}
    \label{fig:jwst_c3_data}
\end{figure}

\begin{table*}
    \centering
    \caption{JWST/MIRI observations of Eps Ind A taken for JWST program GO \#5037.}
    \begin{tabular}{lllllll}
        \toprule
        \toprule
        Target & Filter & Readout & Groups/Int & Ints/Exp & Dithers & Total Time [s] \\
        \midrule
        background\_1 & F1140C & FASTR1 & 697 & 25 & 2 & 8364.3 \\ 
        eps Ind A    & F1140C & FASTR1 & 697 & 25 & 1 & 4182.2 \\ 
        DI Tuc    & F1140C & FASTR1 & 378 & 24 & 5 & 10899.4 \\ 
        background\_2 & F1140C & FASTR1 & 378 & 24 & 2 & 4359.8 \\ 
        \botrule
    \end{tabular}
    \label{tab:jwstobservations}
    \tablecomments{background\_1 and background\_2 are located close to eps Ind A and DI Tuc respectively, and were selected to be free of sources in the WISE catalog. All observations were carried out sequentially between 2025-05-09 to 2025-05-10.}
\end{table*}

\section{Data Reduction}
\label{sec:datareduction}

To create science-ready images, we processed the data with \texttt{spaceKLIP}\footnote{\url{https://spaceklip.readthedocs.io/en/latest/}, commit \texttt{a72c167}} \citep{Kammerer2022,Carter2023}, using largely the same method as in \citet{Matthews2024} which in turn follows \citet{Carter2023}. We briefly summarize the key steps here. We started our reductions from the \texttt{*uncal.fits} files (rather than the processed \texttt{*cal.fits} files), since previous works have shown that adjusting pipeline parameters in the \texttt{jwst} pipeline \citep{Bushouse_jwst} steps that produce \texttt{*cal.fits} files can improve the final data reduction \citep{Carter2023}. In the \texttt{*uncal.fits} files, raw telemetry data has been converted into fits header keywords but no other processing has been performed.

We ran \texttt{step1} of the pipeline using an increased jump rejection threshold of 8 so as to avoid flagging too many pixels as bad. For \texttt{step2} we used standard pipeline settings. We then cleaned badpixels from our images, using several techniques sequentially to remove them: we used the data quality flags, spatially and temporally searched for outliers, and sigma clipped the data, as well as manually flagging and removing badpixels that were not detected by the pipeline. We also subtracted the background, following the custom approach described in \citet{Godoy2024}, and masked some bright pixels in the core of the stellar PSF. Finally, we padded the frames and replaced NaN values within the images. Plate scale, orientation angle, and distortion were not derived independently, and we instead used the calibrations provided as default by the \texttt{jwst} pipeline \citep{Bushouse_jwst} and the corresponding values in the FITS headers (PIXAR\_A2, ROLL\_REF, V3IYANG).

To remove the stellar PSF, we used the \texttt{pyKLIP} package \citep{Wang2015} as implemented within \texttt{spaceKLIP}. This package uses KLIP \citep{Soummer2012}, a principal component analysis based method which builds a stellar PSF model empirically from several reference images, that can then be subtracted from science observations of interest. Our observing sequence included reference star images but not rotated images of the same target, so we carried out an RDI-only reduction, making use of all five dithered observations of the reference star. We explored various combinations of annuli, subsections and number of modes to remove starlight; in Figure \ref{fig:jwst_c3_data} we show an image of the target using a single annulus and subsection, and with 50 Karhunen–Loève modes removed.

\subsection{Photometry of Eps Ind Ab}
\label{sec:photometry}

For a companion as bright as Eps Ind Ab, photon noise is not the limiting uncertainty in measuring the companion photometry. Instead, systematics and calibration terms are important in driving the photometric uncertainty -- and these can be difficult to determine, especially for coronagraphic observations (see e.g.~\citealt{Boccaletti2023,Malin2024}). We therefore take care to derive photometry that is simultaneously (1) accurate and (2) has realistic uncertainties. We follow the approach of \citet{Boccaletti2023}: we use several different methods to extract the photometry, and analyze the spread of results. Our adopted flux is the mean and standard deviation of the value measured with each method. This provides robust values and uncertainties, and additionally provides an independent verification of the host star magnitudes derived in the JWST filters in \citet[][see their Methods section]{Matthews2024} since only some methods take the host star photometry into account. We note however that some potential sources of systematics (e.g.~on the stellar magnitude) would bias all measurements with a certain method equally, that is, they impact the measured apparent magnitude but not the measured color. For our final color determination, we therefore use the mean and standard deviation of the color measured with each method (rather than calculating the color from the adopted apparent magnitudes). For this work, we use three methods to measure magnitudes and colors: (1) \texttt{spaceKLIP} companion injection; (2) a simulation based on \texttt{STPSF}\footnote{\url{https://stpsf.readthedocs.io/en/latest/}} (a tool that produces custom PSF models for JWST for a specified input spectrum and telescope set-up, \citealt{Perrin2014}), but implemented outside of \texttt{spaceKLIP} and run on a data reduction performed entirely independently of \texttt{spaceKLIP}, following \citet{Malin2024}; (3) a method based on a contrast measured relative to the stellar flux and using target acquisition observations, as described in \citet{Boccaletti2023}.
We also update the photometry at F1065C and F1550C originally derived in \citet{Matthews2024}, following the same methods, that is, combining the \texttt{spaceKLIP} photometry (as measured in \citealt{Matthews2024}) with newly derived values using the two additional methods so that the uncertainties are more conservative. 

We provide the final photometric measurements and uncertainties for all filters in Table \ref{tab:photometry}, where we additionally provide photometric measurements derived using each of the individual methods above. The adopted colors of the companion are F1065C-F1140C $=0.88\pm0.08$\,mag and F1065C-F1550C $=1.92\pm0.08$\,mag.

\begin{table*}
    \centering
    \caption{Photometric measurements of Eps Ind Ab in the JWST/MIRI coronagraphic filters.}
    \begin{tabular}{cccc}
        \toprule
        \toprule
          & F1065C & F1140C & F1550C \\
        \midrule
        Flux method 1: \texttt{spaceKlip} & 5.26$\pm$0.07 & 8.75$\pm$0.10 & 6.69$\pm$0.03 \\
        Flux method 2: \texttt{STPSF} & 5.06$\pm$0.06 & 9.98$\pm$0.05 & 7.50$\pm$0.002 \\
        Flux method 3: TA contrast & 4.96$\pm$0.06 & 8.50$\pm$0.05 & 6.22$\pm$0.001 \\
        \midrule
        Adopted Flux & 5.09$\pm$0.12 & 9.08$\pm$0.64 & 6.80$\pm$0.52 \\
        Apparent Magnitude [mag] & 13.13$\pm$0.03 & 12.21$\pm$0.08 & 11.16$\pm$0.08 \\
        Absolute Magnitude [mag] & 15.33$\pm$0.03 & 14.40$\pm$0.08 & 13.36$\pm$0.08 \\
        \botrule
        
    \end{tabular}
    \label{tab:photometry}
\tablecomments{All listed fluxes are in units of $10^{-18}$W/m$^2$/\micron. The adopted fluxes and corresponding uncertainties are the mean and standard deviation of the color measured with each method (see Section \ref{sec:photometry} for details).}
\end{table*}

\section{Detection of Ammonia}
\label{sec:ammonia}

\begin{figure}
    \centering
    \includegraphics[width=0.5\textwidth]{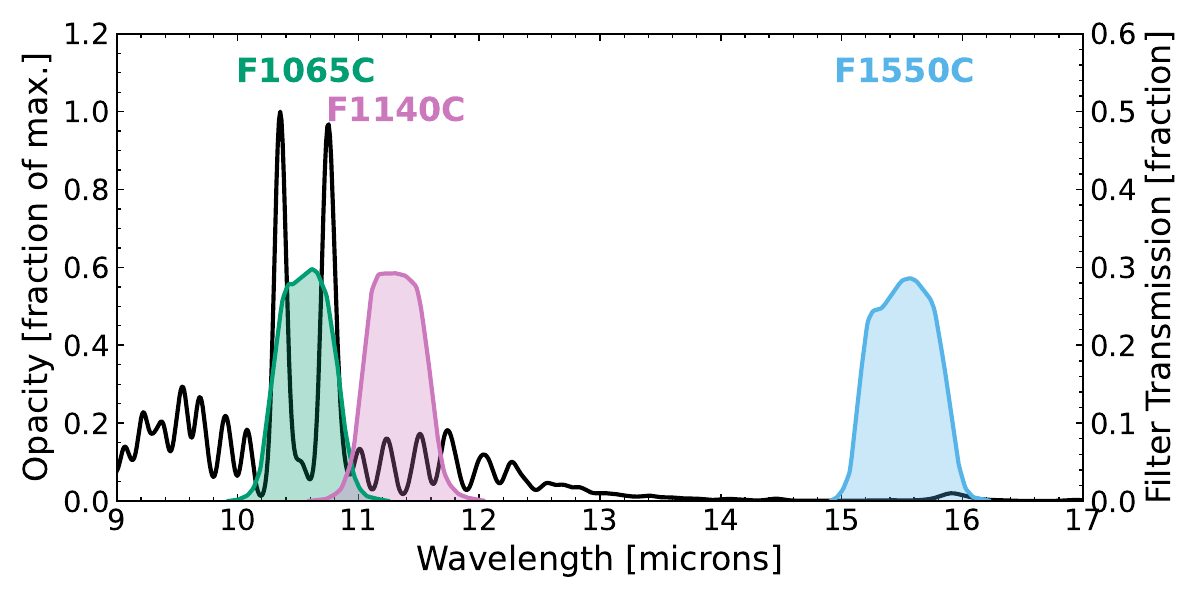}
    \caption{Opacity of ammonia as a function of wavelength (left axis), plotted against the transmission of the MIRI coronagraphic filters (right axis). The deep, double-peaked 10-11\micron~ammonia absorption feature falls primariy in the F1065C filter, and the F1065C-F1140C color is a proxy for the depth of the feature.
    }
    \label{fig:ammonia_opacity}
\end{figure}

The F1065C-F1140C color of Eps~Ind~Ab is $0.88\pm0.08$\,mag, i.e., the planet has an $11\sigma$ difference in apparent magnitude between 10.6\micron~and 11.3\micron. This significant planet brightness difference between observations in two neighboring filters is strong evidence for the presence of an ammonia absorption feature. Figure \ref{fig:ammonia_opacity} shows the filter profiles for the three MIRI coronagraphic filters in which Eps~Ind~Ab has been observed, alongside the opacity as a function of wavelength of ammonia (from \citealt{Coles2019}). Ammonia absorbs strongly within the F1065C filter, but only weakly within the F1140C filter; consequently, a planet with a deep ammonia feature would show a significant difference in apparent magnitude between these two wavelengths.

\begin{figure*}
    \centering
    \includegraphics[width=0.95\textwidth]{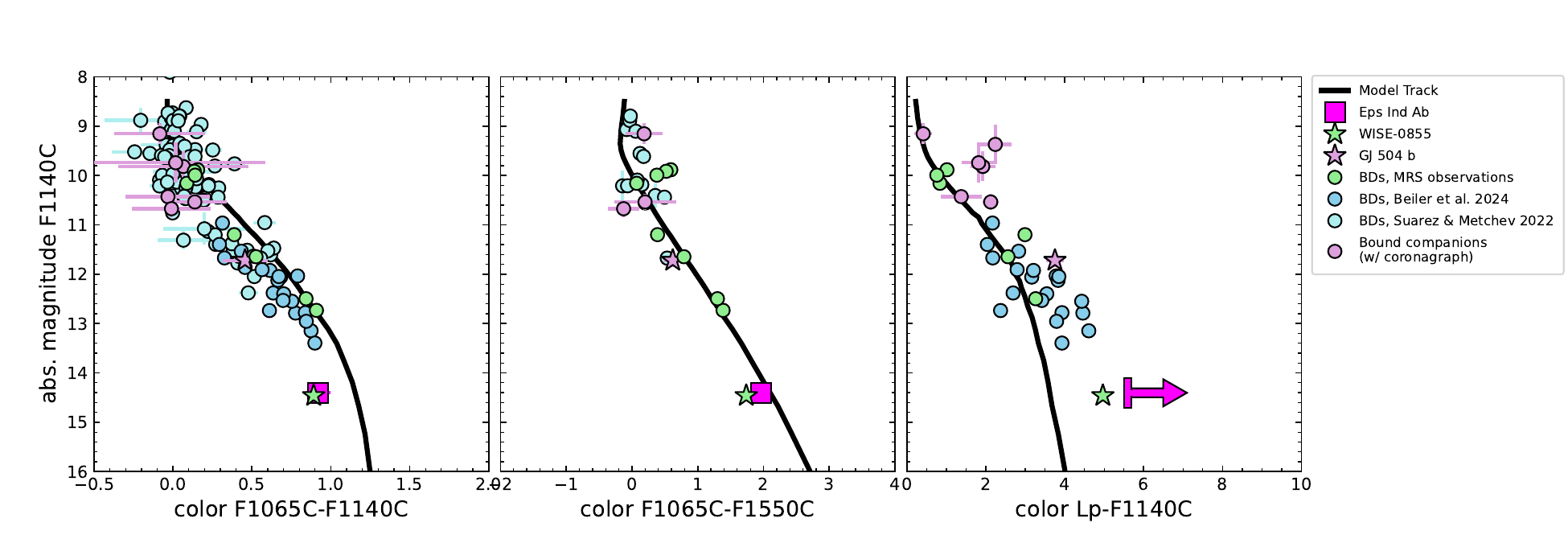}
    \caption{CMD positions of Eps Ind Ab, as well as a number of other cold brown dwarfs (with details provided in Section \ref{sec:archivalbds}), considering the three JWST filters as well as the NaCo L' filter (for which there is a non-detection of Eps Ind Ab; the base of the arrow is the 5$\sigma$ lower limit in L'-F1065C color for this planet). Models are from the Sonora Flame Skimmer grid (Mang et al.~in prep.), and have solar metallicity and C/O and $K_{\rm zz}=10^7$\,cm$^2$/s. These models broadly explain the population of warmer brown dwarfs, and in particular the trend whereby the F1065C-F1140C color becomes increasingly red, tracing an increasingly deep ammonia feature, in cooler brown dwarfs. However, both Eps Ind Ab and WISE 0855 show significantly bluer F1065C-F1140C colors than these simple models, indicating that the ammonia feature is smaller than expected. Eps Ind Ab and WISE 0855 have remarkably similar mid-IR magnitudes and colors, but Eps Ind Ab is significantly redder in L'-F1065C (i.e., it is significantly fainter at L'), and we discuss potential explanations in the text.}
    \label{fig:ammonia_cmd}
\end{figure*}

To contextualize the measured photometry of Eps~Ind~Ab, we plot its position on color-magnitude diagrams (CMDs) in Figure \ref{fig:ammonia_cmd}, alongside the positions of a ``default" atmosphere model from the Sonora Flame Skimmer model set (Mang et al.~in prep.)~with solar metallicity and carbon-to-oxygen ratio (C/O). Sonora Flame Skimmer is the next set of clear atmospheric and evolutionary models that extends the Sonora Bobcat \citep{Marley2021} and Sonora Elf Owl \citep{Mukherjee2024, Wogan2025} models to colder temperatures, a wider range of metallicities, and lower surface gravities. The Sonora Flame Skimmer grid spans $T_{\rm eff}$ = 50-2400 K, $\log(g)$ = 2.0-5.5, [M/H] = -1.0-+2.0, and C/O = 0.5-2.5 (relative to solar, where solar = 0.458). The models include both equilibrium and disequilibrium chemistry with vertical mixing parameters of $K_{\rm zz}$ = 10$^{2}$–10$^{9}$ cm$^{2}$ s$^{-1}$. While the equilibrium chemistry models have always included rainout processes, the disequilibrium models in Sonora Elf Owl did not, as the abundances of quenched species were held constant above the quench point. For colder atmospheres, however, volatile species such as H$_2$O, CH$_4$, and NH$_3$ are expected to condense and rain out. In Sonora Flame Skimmer, we adopt a hybrid treatment of disequilibrium chemistry in which these species continue to follow their saturation vapor pressure curves above the quench point. Sonora Flame Skimmer also incorporates the improved CO$_2$ chemistry treatment introduced in Sonora Elf Owl v2 \citep{Wogan2025}.

The CMDs are constructed based on the JWST coronagraphic filters, as well as the L' filter (at 3.8\micron, where Eps~Ind~Ab has not been detected but for which there is a deep upper limit on the exoplanet flux based on NaCo observations, \citealt{Viswanath2021}). We also show observational data for other cold exoplanets and brown dwarfs, with details on how these were selected and relevant references listed in Appendix \ref{sec:archivalbds}. Warm substellar objects (absolute F1065C magnitude $\gtrsim$10, corresponding to temperature $\gtrsim$1500\,K) have an F1065C-F1140C color close to 0. %
Below this temperature, objects are increasingly red in F1065C-F1140C color, due to an increasingly deep ammonia absorption feature. GJ504b, with its F1065C-F1140C color of $0.45\pm0.14$\,mag, is the only planet to date where a detection of ammonia has been claimed based on MIRI photometric observations; Eps~Ind~Ab is significantly redder than GJ504b in these filters with a color of $0.88\pm0.08$\,mag.

Eps~Ind~Ab is bluer than the ``default'' models predict at this temperature, and bluer than all of the models presented in \citet{Matthews2024} in F1065C-F1140C color. Strikingly, the cold brown dwarf WISE 0855 \citep[$\sim$285\,K,][]{Luhman2014,Kuhnle2025} has almost the same F1065C-F1140C color as Eps~Ind~Ab, and indeed \citet{Kuhnle2025} found the ammonia abundance of WISE 0855 to be less than expected from chemical equilibrium modelling, based on their retrieval analysis. In other words, both of these very cold objects show shallower ammonia features than expected, while all of the warmer brown dwarfs shown in Figure \ref{fig:ammonia_cmd} are in broad agreement with the models. The consistency in F1065C-F1140C color of Eps~Ind~Ab and WISE 0855 suggests that this shallower ammonia feature is typical of very cold atmospheres. The physical mechanism producing this shallower-than-expected ammonia feature is not yet clear, so we now explore several possible scenarios. 

\begin{figure*}
    \centering
    \includegraphics[width=0.95\textwidth]{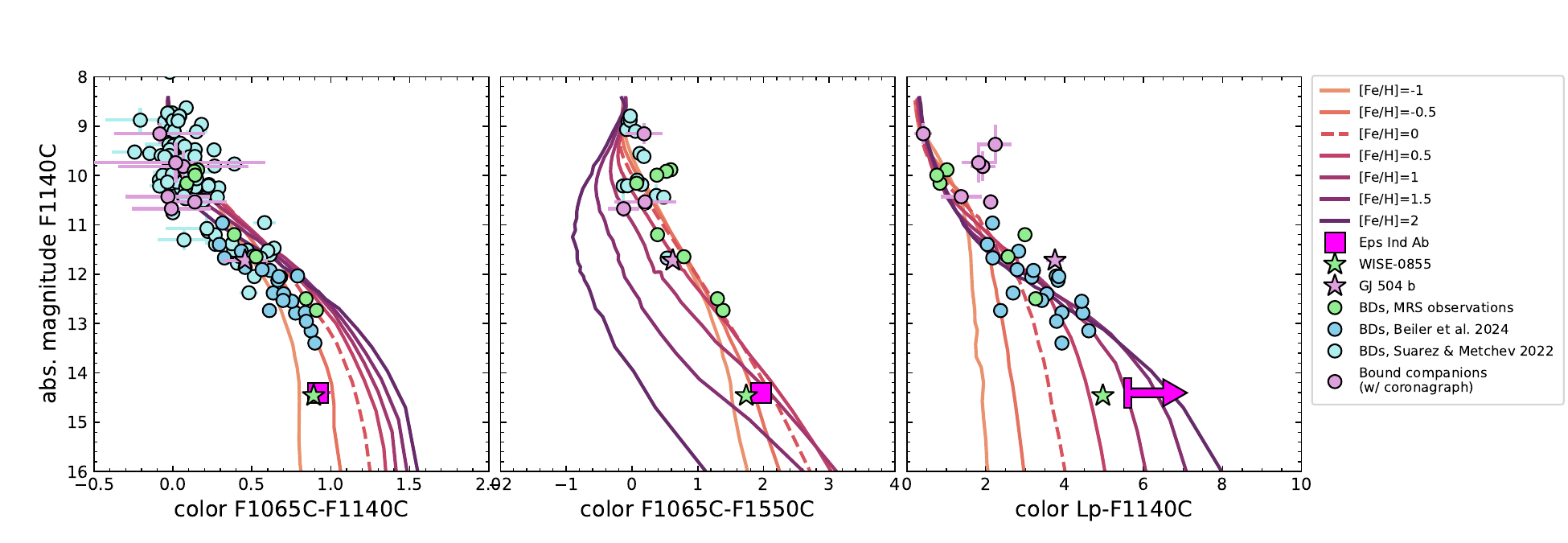}
    \caption{Same as Figure \ref{fig:ammonia_cmd}, but for models with a range of metallicities. While a significantly sub-solar metallicity would explain the shallow ammonia feature of Eps Ind Ab (left), the models with the lowest metallicity have the brightest NaCo L' magnitude. Sub-solar metallicity is not (by itself) sufficient to explain the in-hand photometric constraints for Eps Ind Ab. All plotted metals have solar C/O and $K_{\rm zz}=10^7$\,cm$^2$/s, and the model with [Fe/H]=0 (identical to Figure \ref{fig:ammonia_cmd}) is highlighted with a dashed line.}
\label{fig:ammonia_cmd_metallicity}
\end{figure*}

A small ammonia feature could indicate a low metallicity for Eps~Ind~Ab: in this scenario, molecular abundances in the atmosphere would be lower, and their absorption features would be suppressed. The host star Eps~Ind~A has a slightly sub-solar metallicity \citep[$-0.06\pm0.08$][]{Santos2004}, and in a gravitational instability formation pathway this could have been inherited by the planet. Metal-poor Sonora Flame Skimmer models with [M/H] = -1.0 show a good match to the F1065C-F1140C color of Eps~Ind~Ab (see Figure \ref{fig:ammonia_cmd_metallicity}). However, this explanation for the shallow ammonia feature is unsatisfactory: at low metallicity models predict an increased near-IR (3-5\micron) brightness for the planet, since there is less atmospheric absorption from molecules such as \ce{CO}, \ce{CO2} and \ce{H2O}. This is inconsistent with archival non-detections of Eps Ind Ab at these wavelengths \citep{Janson2009,Viswanath2021}, and indeed \citet{Matthews2024} argued that Eps Ind Ab likely has an \textit{elevated} metallicity based on its 3-5\micron~faintness. 

We also explored the impact of the vertical mixing (parameterized with the eddy diffusion coefficient, $K_{\rm zz}$). We tested values between $K_{\rm zz}=10^{2}$\,cm$^2$/s to $K_{\rm zz}=10^9$\,cm$^2$/s, but found that the vertical mixing has no impact on the F1065C-F1140C color in our models, and thus cannot be responsible for suppressing the ammonia feature.

\begin{figure*}
    \centering
    \includegraphics[width=0.95\textwidth]{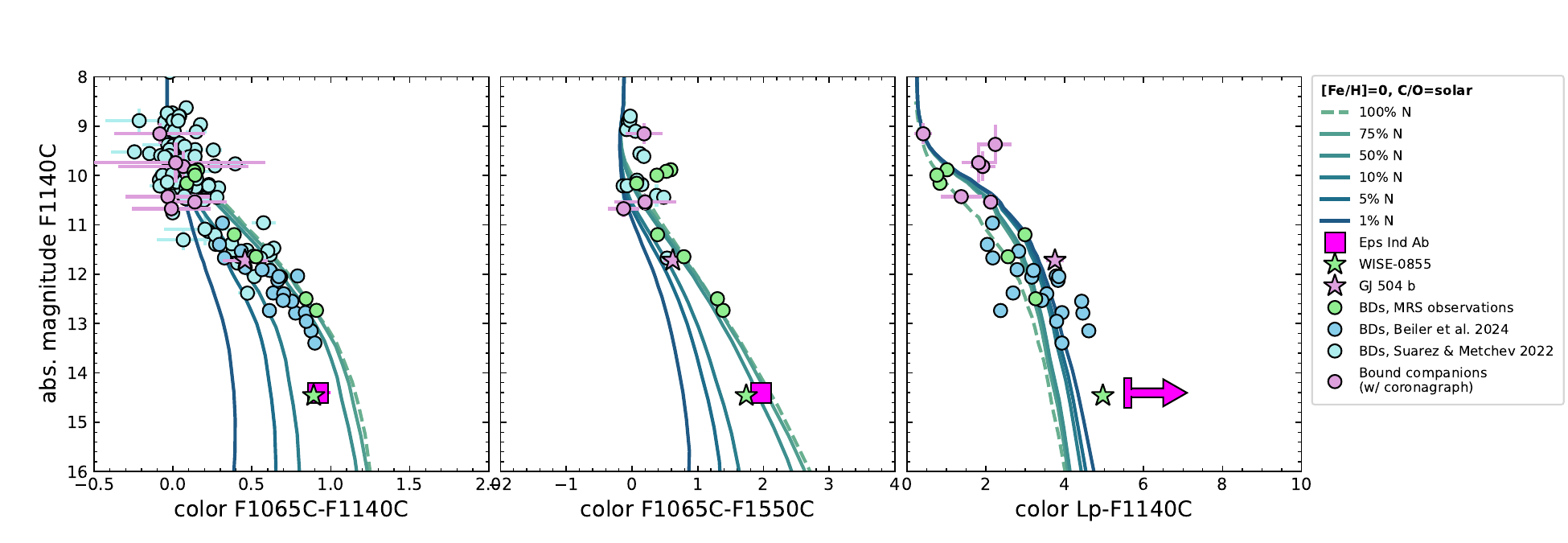}
    \includegraphics[width=0.95\textwidth]{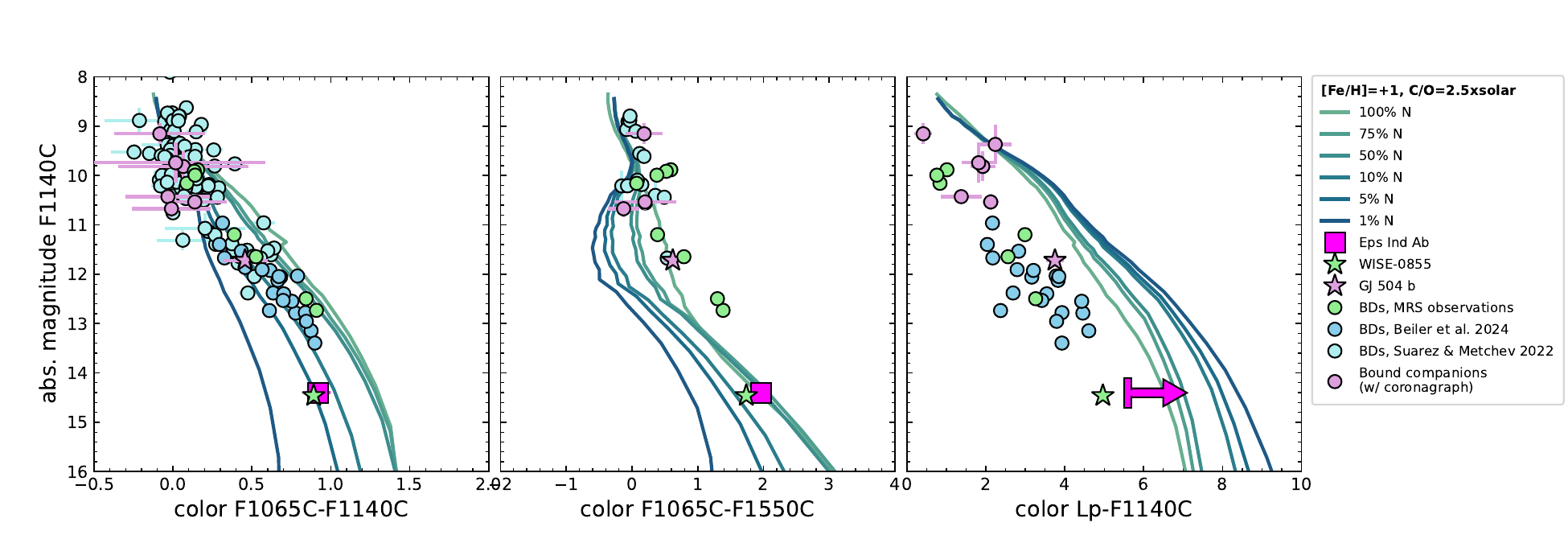}
    \caption{Same as Figure \ref{fig:ammonia_cmd}, but for models depleted in nitrogen. The top and bottom rows show respectively the case of solar metallicity and C/O, and of enhanced metallicity and C/O as found for Eps~Ind~Ab in \citealt{Matthews2024}. Percentage labels indicate the nitrogen content of each atmosphere as a fraction of equilibrium models, and the model from Fig.~\ref{fig:ammonia_cmd} is highlighted with a dashed line. Atmospheres strongly depleted (to 5\%) in nitrogen are consistent with both the F1065C-F1140C color of Eps Ind Ab and WISE 0855, and strongly N-depleted atmospheres also have bluer L'-F1065C colors. However, only in models with \textit{both} (1) depleted nitrogen and (2) enhanced metallicity and C/O is the L'-F1065C color sufficiently blue to match the observational constraints for Eps Ind Ab.}
    \label{fig:ammonia_cmd_depletion}
\end{figure*}

Another explanation to consider for the small ammonia feature is that nitrogen alone is depleted in the exoplanet atmosphere (Figure \ref{fig:ammonia_cmd_depletion}). To test this idea we generated a version of the Sonora Flame Skimmer models with nitrogen depletion but otherwise identical to the default grid models. In Figure \ref{fig:ammonia_cmd_depletion} we show these models, with either solar metallicity and C/O or the elevated metallicity and C/O inferred by \citet{Matthews2024}. An atmosphere with $\sim5-15\%$ of the equilibrium nitrogen abundance (a depletion of $\sim85-95\%$ of the nitrogen) shows close agreement with the measured F1065C-F1140C color. Removing nitrogen from the atmosphere also increases the L'-F1065C color in these models, and therefore may provide an explanation for the archival non-detections of Eps Ind Ab in the 3-5\micron~range \citep{Janson2009,Viswanath2021}, though we note that nitrogen depletion alone is not sufficient to explain the Eps Ind Ab L' upper limit and some enrichment of C/O and metallicity is also required. Nonetheless, Figure \ref{fig:ammonia_cmd_depletion} demonstrates that with significant nitrogen depletion, and with fine-tuning of the carbon and oxygen atom fractions in a cold atmosphere, the CMD positions of both Eps Ind Ab and WISE 0855 can be explained.

With this in mind, we briefly consider possible explanations for an atmosphere that is depleted in nitrogen (but not carbon or oxygen). There are two explanations worth considering: perhaps this is a signature of formation in a disk, or perhaps nitrogen is actively being removed from the atmosphere through cloud formation and/or rain-out processes. However, neither is a perfect explanation for our observations. Considering the first, it is commonly thought that the formation mechanism and location of a planet within its disk should impact its composition \citep[e.g.][]{Oberg2011}, and \citet{Turrini2021} modeled the impact of formation location on nitrogen content of the atmosphere. That work found that some formation pathways can lead to planets depleted in nitrogen, but none of their scenarios had depletion as strong as we infer for Eps Ind Ab (85-95\%). Further, this formation hypothesis would only apply in the case of WISE 0855 if that brown dwarf is in fact an ejected planet. The second possibility is that nitrogen atoms are removed from the atmosphere through the formation of clouds. Ammonia-ice clouds are expected to form in very cold atmospheres \citep[e.g.][]{Sudarsky2000,Sudarsky2003}; these are also observed in Jupiter \citep[e.g.][]{Irwin1998}. However, in our model with water clouds, \ce{NH3} never reaches the pressure-temperature regime required for \ce{NH3} to condense. We note that the pressure-temperature profile for a model without water clouds does cross the \ce{NH3} condensation line (see Appendix \ref{sec:condensation}), but there is no reason to expect that such cold atmospheres could appear without water-ice clouds. Another cloud species that may contribute to the nitrogen depletion is ammonium hydrosulfide \citep[NH$_4$SH;][]{Visscher2006}. This cloud forms between the NH$_3$ and H$_2$O cloud decks through a reaction between NH$_3$ and H$_2$S, and is observed in Jupiter \citep{Irwin1998}. However, its formation is limited by the abundance of H$_2$S, which is expected to be on the order 4 times less abundant than NH$_3$, and therefore cannot independently account for the amount of nitrogen that would need to be depleted to explain the Eps~Ind~Ab color. The final possible avenue for nitrogen depletion is the dissolution of NH$_3$ into H$_2$O clouds to form a mixed solution. This process has been investigated in the context of NH$_3$ depletion in Jupiter \citep{Guillot2020a, Guillot2020b}, where vigorous mixing of H$_2$O ice particles into higher altitudes could allow the NH$_3$ to dissolve and produce an aqueous solution. Further modeling is necessary to evaluate the effects of this mechanism in colder exoplanetary atmospheres, and future works should also search for absorption signatures of these various cloud species in the Eps~Ind~Ab atmosphere: \ce{NH3} clouds, in particular, should have several strong features in the near- and mid-IR \citep{Martonchik1984}.

\begin{figure*}
    \centering
    \includegraphics[width=0.95\textwidth]{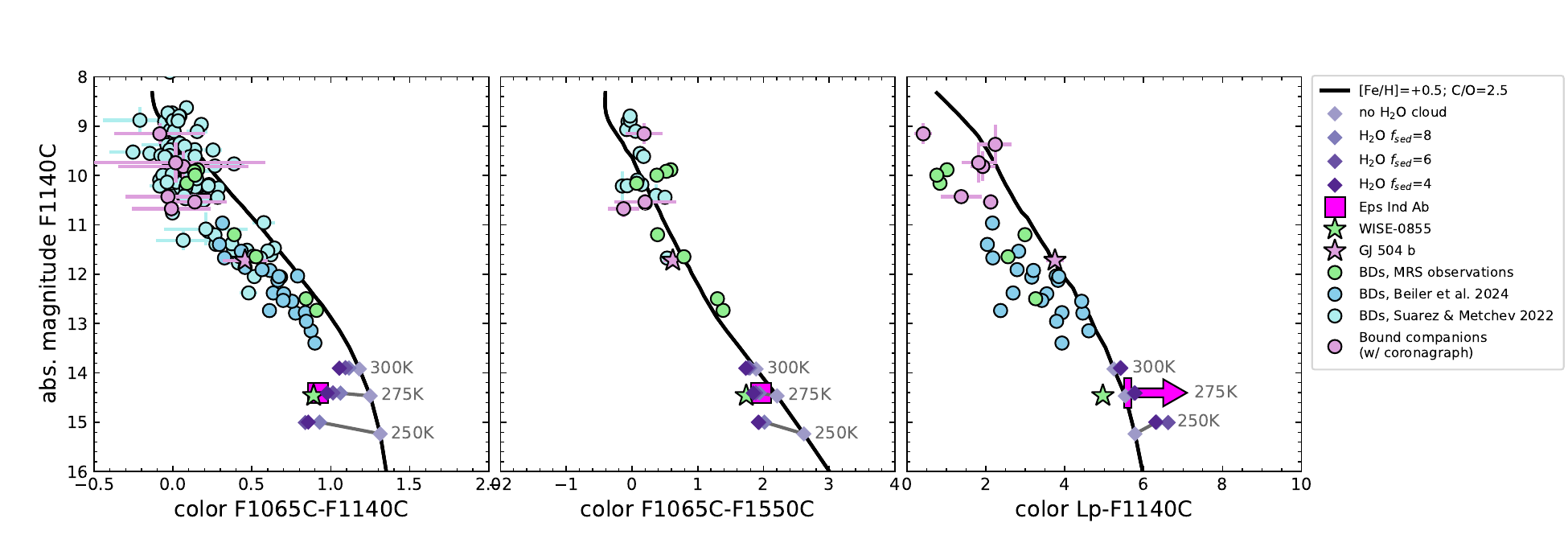}
    \includegraphics[width=0.95\textwidth]{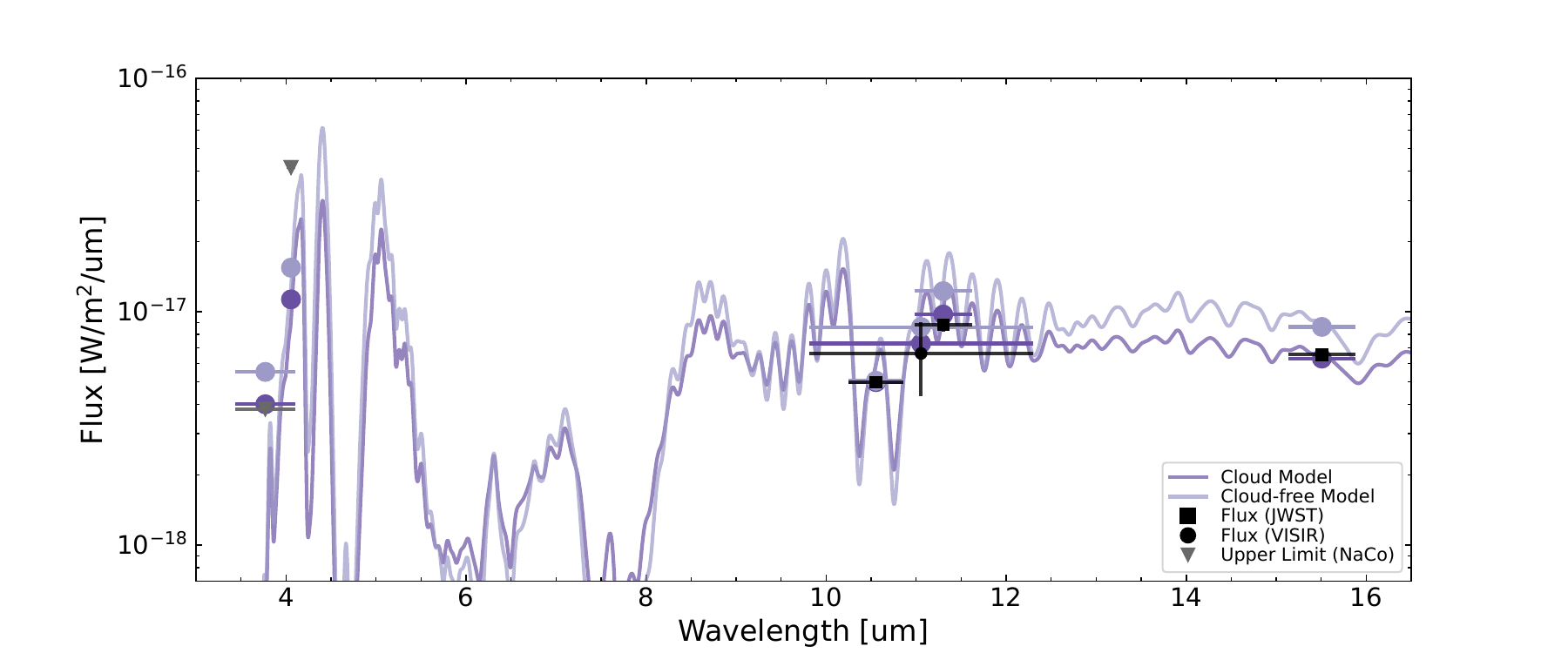}
    \caption{\textbf{Upper panel:} CMD as in Fig.~\ref{fig:ammonia_cmd}, but with the impact of clouds highlighted. The black line shows the trend for the best-fit [Fe/H] and C/O from our small grid (see text), while the purple diamonds show the impact of introducing increasingly thick water-ice clouds (parametrized via $f_{\rm sed}$), for models with temperatures 250\,K, 275\,K and 300\,K. Clouds have a more dramatic impact on the photometry at lower temperatures where they reduce the relative depth of the ammonia feature, making the F1065C-F1140C and F1065C-F1550C colors redder, and subdue the near-IR emission, making the Lp-F1140C color redder.} \textbf{Lower panel:} Model emission spectrum for an atmosphere with thick water-ice clouds, generated with PICASO (\citealt{Batalha2019,Mukherjee2023,Mang2026}), and described in detail in the text. We also show a model without water-ice clouds but otherwise identical for reference. The water-ice cloud model is a best-fit from a small grid of models, and is compatible with all in-hand photometry and upper limits for Eps Ind Ab. 
    \label{fig:cloud_model}
\end{figure*}

Finally, we turn to our preferred explanation for the small ammonia feature: perhaps this indicates thick water-ice clouds in the Eps~Ind~Ab atmosphere. Such clouds could have the dual impacts of (1) reducing the depth of the ammonia feature, thereby producing a bluer F1065C-F1140C color and (2) suppressing the near- and mid-IR emission of the exoplanet, consistent with archival non-detections of this object from the ground between 3 to 5\micron{} \citep{Janson2009, Viswanath2021}. To test this, we generated a mini-grid of custom models with PICASO \citep{Batalha2019,Mukherjee2023,Mang2026} that self-consistently treat clouds using \texttt{Virga} \citep{virga_code, virga_paper1, virga_paper2}, a Python-based implementation of the \citet{Ackerman2001} cloud model. This grid spans $T_{\rm eff}$ = [250, 275, 300~K], log(g) = [4.0, 4.25, 4.5, 4.75] (cgs), [M/H] = [0, +0.5, +1.0], and C/O = [1.0, 2.5]. For the disequilibrium models, we parametrize the strength of vertical mixing using an eddy diffusion coefficient of $K_{\rm zz} = [10^2,10^9]$\,cm$^2$/s. H$_2$O clouds are parameterized by the sedimentation efficiency parameter, $f_{\rm sed}$, where larger values produce optically thinner, vertically constrained clouds with large particles, and smaller $f_{\rm sed}$ values result in optically thicker, more vertically extended clouds with smaller particles. In Figure \ref{fig:cloud_model} (upper) we demonstrate the effect of introducing water-ice clouds into a model atmosphere, at each of the tested temperatures, using a model with elevated metallicity and C/O as a baseline. Water-ice clouds have more impact at colder temperatures, and produce the expected bluer F1065C-F1140C and redder Lp-1140C colors. We also show an example of a custom PICASO model that includes H$_2$O clouds self-consistently, as well as a similar model without water-ice clouds, in Figure \ref{fig:cloud_model} (lower). This is the best-fit spectrum from our grid, and is a 275~K object with $\log{g} = 4.5$, $3\times$ solar metallicity, $2.5\times$ solar C/O, $K_{\rm zz} = 10^9$\,cm$^2$/s, and $f_{\rm sed} = 6$. The model contains a highly optically thick H$_2$O cloud with a column optical depth of 416, reaching $\tau = 1$ near 0.7\,bar. Atmospheric models including water-ice clouds can explain the colors of Eps Ind Ab: the model matches the three JWST photometric points, and suppresses the NaCo L' flux of Eps~Ind~Ab to below the NaCo upper limit (\citealt{Viswanath2021}, but we use the recalculated NaCo upper limit from \citealt{Matthews2024} for Figure \ref{fig:cloud_model}). Customized cloud models for both Eps Ind Ab and WISE 0855 could explain why both objects have suppressed ammonia features but one is fainter than the other in the 3-5$\micron$ region: for example, perhaps Eps Ind Ab has uniform cloud cover while WISE 0855 has only patchy water-ice clouds with holes allowing the near-IR flux to escape, or both objects have similar water-ice clouds but different underlying metallicities and/or carbon-to-oxygen ratios.

While a model atmosphere with water-ice clouds, elevated metallicity, and elevated C/O ratio can explain all in-hand photometry, questions remain. Water-ice clouds are naturally expected in this cold atmosphere, but the elevated metallicity and C/O of our best-fit remains a challenge for formation models. Future studies that expand the wavelength coverage of Eps Ind Ab could more definitively confirm whether or not there are water-ice clouds in this atmosphere, and in particular spectroscopic measurements could provide a more precise measure of the metallicity and C/O ratio, which would be valuable for studying the formation of this cold giant planet, and the cloud $f_{\rm sed}$, which would be valuable in further constraining the onset of water-ice clouds in cold atmospheres and their typical properties.

\section{Confirmation of common proper motion}
\label{sec:cpm}

The most definitive evidence that a point source is a planet -- and not a background star or galaxy -- is the detection of common proper motion between the star and the point source. The significant proper motion of Eps Ind A makes confirming proper motion trivial with more than one epoch of observations: as one of the closest stars to us, Eps Ind A has a proper motion of 4708\,mas/year \citep{gaia_edr3}. Thus, our F1140C re-detection of a point source in the NW quadrant trivially confirms common proper motion between the host star and the point source: a background object would be displaced by [7.57, -4.62] arcsec ([68.8, 42.0] pixels) between the 2023 and 2025 JWST observations of the system, inconsistent with our re-detection of the planet in the NE quadrant of the 2025 images. \citet{Matthews2024} previously argued that Eps~Ind~Ab was a planet, based in part on archival VISIR/NEAR data collected in 2019 (see also \citealt{Pathak2021}). That dataset showed a detection of a point source with signal-to-noise $\sim3$ in the upper left quadrant, consistent with the much more robust point source detection in the 2023 JWST observations. The current dataset (2025 JWST observations) provides the second epoch of observations with $>5\sigma$ detection significance, and the third detection at any significance of the companion. This result is a definitive demonstration that the planet shares common proper motion with its host star, and as a consequence also confirms the conclusion of \citet{Matthews2024} that the low-significance VISIR detection corresponds to the same point source as the JWST detection.

\section{Updated orbit fits}
\label{sec:orbits}

With our new image astrometry in hand, we re-fit the orbit of Eps~Ind~Ab and derive an updated dynamical mass. For this orbit fit we follow largely the same method as described in \citet{Matthews2024}, with a few key changes. Similar to that work we used \texttt{orvara} \citep{BrandtOrvara} to fit (1) the direct imaging astrometry from this work and from \citet{Matthews2024}; (2) the Hipparcos-Gaia acceleration (also referred to as the Proper Motion Anomaly) of the host star \citep{Brandt2018,Kervella2019}, using the acceleration derived in the \citet{Brandt2021} catalog which updates \citet{Brandt2018} with the inclusion of \textit{Gaia} EDR3 data; (3) RV data from the CES long camera and very long camera \citep[LC and VLC respectively;][]{Zechmeister2013}, from UVES \citep{Feng2025}, and from HARPS using the data reduction presented in \citet{Trifonov2020} and split into three separate regimes where different instrument offsets are expected (refered to as HARPSpre, HARPSpost and HARPSpost2 in \citealt{Feng2019}). The changes in our orbit fitting setup relative to \citet{Matthews2024} are as follows:

\begin{itemize}
    \item We use one additional direct imaging astrometry point, derived from the new JWST data presented in this work. The planet has separation $3.654\pm0.018$\,arcsec and position angle $35.065\pm0.036^\circ$ as measured with JWST on 2025-05-10.
    \item We correct the LC and VLC radial velocity data for perspective/secular acceleration, as also described in \citet{Feng2025}. This effect describes the change in the measured radial velocity of a target, due to our changing viewing angle as the target moves across the sky (as opposed to an inherent change in intrinsic velocity of the target), and is particularly relevant for the highest proper-motion stars in the sky, such as Eps Ind A \citep{Kuerster2003}. Note that for HARPS and UVES, published data already include a perspective acceleration correction, while for LC and VLC the published data do not include this correction. 
    \item We apply a nightly binning to all datasets, whereas \citet{Matthews2024} applied nightly binning only to the HARPS data. Thus, our fit includes 266 independent RV points, with 202 from HARPS, 44 from LC, 17 from VLC, and 3 from UVES. For nights with 3 or fewer measurements from a single instrument we combined RV uncertainties in quadrature, while for nights with more than 3 measurements we took the standard deviation of the unbinned RV measurements as the RV uncertainty. %
\end{itemize}

\begin{figure*}
    \centering
    \includegraphics[width=0.95\textwidth]{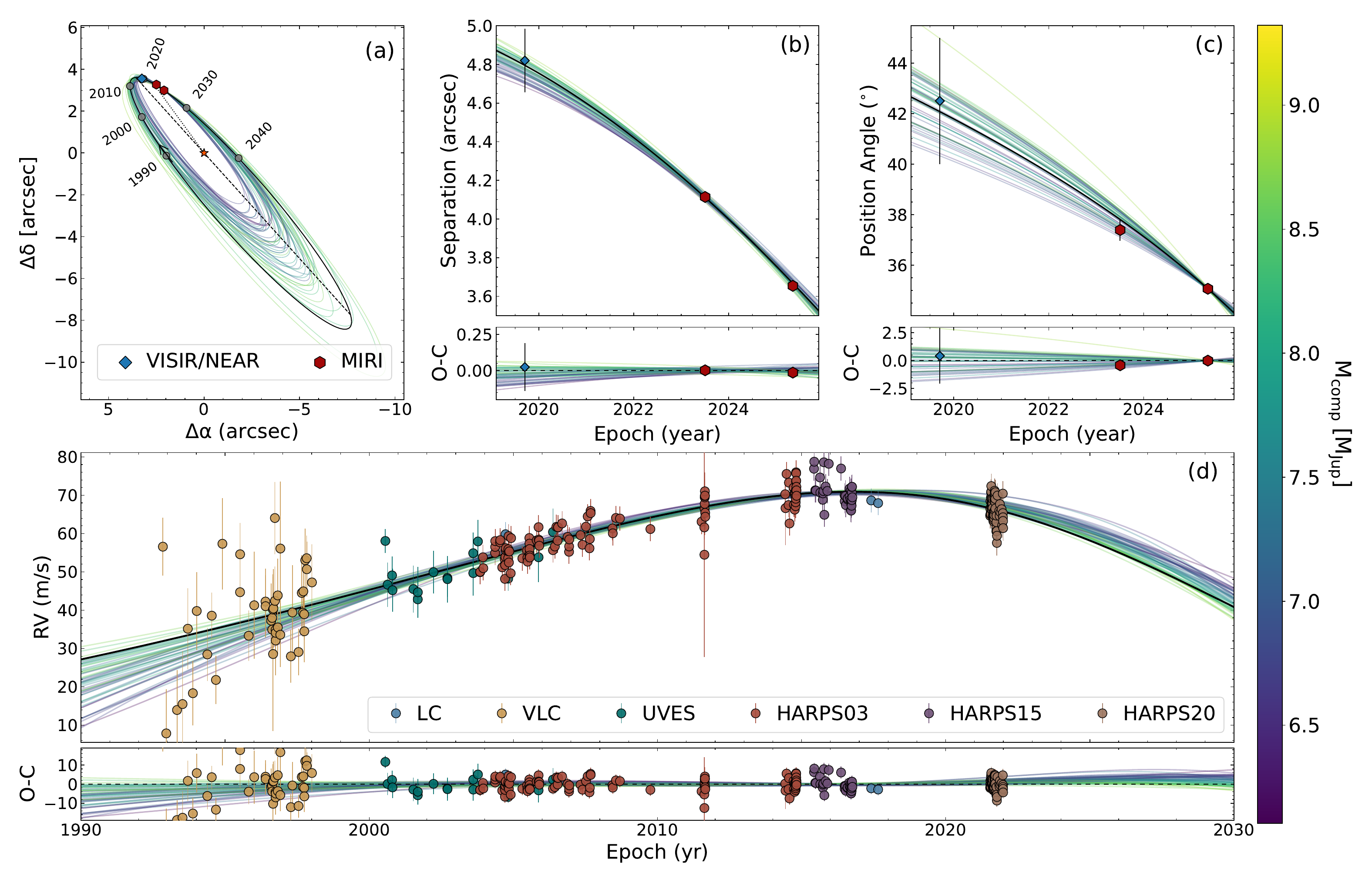}
    \caption{Best-fit orbit solutions. Here we show the on-sky orbit traced by Eps~Ind~Ab in panel (a); the evolution of on-sky separation and position angle of the planet during our observational baseline in panels (b) and (c) respectively; and the RV data in panel (d). The Maximum A Posteriori (MAP) solution is shown with a thick black line, alongside 50 randomly drawn samples from the posterior, with the same colorbar applying to all panels. Our measured JWST/MIRI and VISIR/NEAR astrometry is shown with red hexagons and a blue diamond respectively, and the predicted on-sky orbit position of the exoplanet between 1990-2040 is indicated with small grey points.}
    \label{fig:orbits}
\end{figure*}

The on-sky orbit and evolution of separation and position angle of the planet relative to its host star are shown in Figure \ref{fig:orbits} for our best-fit orbit. In Appendix \ref{sec:orbit_corner} we include a corner plot of key orbital parameters that highlights the correlations of these parameters and the impact on the orbit fit of the newest astrometric point derived in this work. 

We find a more massive planet (${7.63}_{-0.70}^{+0.73}M_{\text{Jup}}$) than \citet{Matthews2024} with this set-up, with a smaller semi-major axis (${20.9}_{-3.3}^{+5.8}\mathrm{AU}$) and a lower eccentricity (${0.244}_{-0.083}^{+0.11}$). This change is driven primarily by the corrected perspective acceleration treatment of the RV data, and the updated binning of the RV points, rather than the additional image astrometry point (see Appendix \ref{sec:orbit_corner} and \citealt{Feng2025}). The best-fit orbit explains all available RV, astrometric and direct imaging datasets; our mass shows good agreement with the findings of \citet{Feng2025} but we find an orbital period that is $\sim1.2\sigma$ lower than in \citet{Feng2025} and an eccentricity that is $\sim1.9\sigma$ lower than in \citet{Feng2025}. The semi-major axis and eccentricity are strongly correlated for an orbit with incomplete phase coverage; Eps Ind Ab is in this regime with $\sim$30 years of RV data and a best-fit period of $\sim$110 years. Thus, it is unsurprising that both the semi-major axis and eccentricity diverge in the same direction from the values of \citet{Feng2025}. Note also that with the orbit fit in this work, orbits close to circular are compatible with the data, and would have semi-major axis $\sim$19\,au.

Considering the on-sky orbit of Eps~Ind~Ab, the new astrometric point provides a significant improvement in the inclination constraint: the precision is doubled, and the derived value is $\sim1.3\sigma$ lower than for a fit without the new astrometry included. This is unsurprising, given that we now have two precise astrometric measurements of the planet with JWST; the updated inclination also has a small impact on the derived mass (since the radial velocity data is sensitive to the $m\sin{i}$ of the planet, which changes by $\sim2\%$ in the fit with two JWST epochs vs the fit without). Ongoing monitoring of the radial velocity and relative astrometry in this system will remain important for refining the orbital solutions, and in particular, sufficiently long-term monitoring will help to break the degeneracy between eccentricity and semi-major axis (Rajpoot et al.~in prep.). 

\begin{table*}
\centering
\caption{Best-fit parameters for the orbit of Eps~Ind~Ab, listed with 1$\sigma$ uncertainties. }
\bgroup
\def\arraystretch{1.3} 
\begin{tabular}{lll}
\toprule
\toprule
Parameter & Median Posterior & Prior \\
\midrule
\midrule
\multicolumn{3}{l}{\textit{Fitted Parameters}}\\
\midrule
    $M_{\text{host}}[M_{\odot}]$        & ${0.781}_{-0.037}^{+0.036}$    & $0.76 \pm 0.04 M_{\odot}$ (Gaussian)     \\
    $M_{\text{comp}}[M_{\text{Jup}}]$   & ${7.63}_{-0.70}^{+0.73}$       & $1 / M_{\text{comp}}$      \\ 
    $a\;[\mathrm{AU}]$                  & ${20.9}_{-3.3}^{+5.8}$         & $1/a$                       \\
    $\sqrt{e} \sin \omega$              & ${0.383}_{-0.10}^{+0.079}$    & Uniform                                  \\
    $\sqrt{e} \cos \omega$              & ${0.22}_{-0.36}^{+0.25}$     & Uniform                                  \\
    $i\;[^{\circ}]$                     & ${102.3}_{-1.7}^{+1.9}$        & $\sin (i)$, $0^{\circ}<i<180^{\circ}$    \\
    $\Omega\;[^\circ]$                  & ${44.6}_{-1.3}^{+1.3}$  & Uniform                                  \\
    $\lambda_{\text{ref}}\;[^\circ]$    & ${356.1}_{-9.8}^{+8.6}$       & Uniform                       \\
    Parallax (mas)                      & ${274.844}_{-0.015}^{+0.015}$      & $274.84 \pm 0.096$ mas (Gaussian) \\
    RV Jitter $\sigma_{\mathrm{jit}}\;[m/s]$ & ${3.23}_{-0.16}^{+0.17}$  & $1 / \sigma_{\mathrm{jit}}$,  $\sigma_{\mathrm{jit}} \in (0,1000] \;m/s$ \\
\midrule
\midrule

\multicolumn{3}{l}{\textit{Derived Parameters}}\\

\midrule
    $P\;(\mathrm{yr})$      & ${108}_{-25}^{+49}$                & $\cdots$ \\
    e                       & ${0.244}_{-0.083}^{+0.11}$          & $\cdots$ \\
    $\omega\;[^{\circ}]$    & ${62}_{-27}^{+48}$              & $\cdots$ \\
    $a\;[\mathrm{arcsec}]$     & ${5.7}_{-0.9}^{+1.6}$          & $\cdots$ \\
\botrule
\botrule
\end{tabular}
\egroup
\tablecomments{Priors for stellar mass prior and parallax are based on \citet{Demory2009} and \citet{gaia_dr3} respectively.}
\end{table*}
\section{Discussion \& Conclusions}
\label{sec:discussion}

In this work we have presented new coronagraphic observations of the Eps Ind A system, collected at 11.3$\micron$ during the 2nd JWST epoch for this target. Combining the new and archival data, we make the following key conclusions:

\begin{itemize}
    \item We find photometric evidence for the presence of ammonia in the planet Eps Ind Ab, with a shallower absorption feature than expected (F1065C-F1140C=$0.88\pm0.08$\,mag). This could imply a low-metallicity planet (a conclusion which is in tension with the L' non-detection of the planet), or a strongly depleted nitrogen abundance in the atmosphere. However, our preferred explanation is that this indicates the presence of water-ice clouds in the exoplanet atmosphere. Even with clouds, the photometry still suggests an elevated C/O and metallicity for Eps Ind Ab, which remains a challenge for formation models.
    \item We present updated orbit fits for the companion Eps~Ind~Ab, with a somewhat higher mass ($7.6\pm0.7$\Mjup) and lower eccentricity ($0.24^{+0.11}_{-0.08}$) than were derived in \citet{Matthews2024}. This change in orbital parameters is largely driven by an updated treatment of the radial velocities, though the new JWST imaging point is particularly helpful in refining the on-sky inclination of the orbit.
    \item We strongly confirm that Eps~Ind~Ab shares common proper motion with its host star (as already inferred in \citealt{Matthews2024}) with the third detection epoch (and the second at $>5\sigma$) of the companion.
\end{itemize}

Eps~Ind~Ab joins a small but growing population of exoplanets with observed ammonia, namely the directly imaged GJ~504~b \citep{Malin2025}, and possibly the transiting WASP-107b \citep{Welbanks2024} and the directly imaged 51~Eri~b \citep{Whiteford2023}. Eps~Ind~Ab is the coldest planet with an ammonia detection to date. Eps~Ind~Ab and WISE 0855 are the only two substellar objects that both (1) have ammonia detections and (2) are also cold enough for water- and possibly nitrogen-containing cloud species to potentially condense in their atmospheres. Both have a shallow ammonia feature, raising an important open question about the physics driving this ammonia suppression that should be investigated further. Upcoming photometric and spectroscopic JWST observations of Eps~Ind~Ab between 3-20\micron~(GO \#8438, \citealt{Berne_jwstgo8438} and \#8714, \citealt{Xuan_jwstgo8714}) will provide an increasingly detailed look at this cold atmosphere, and in particular should provide a clearer answer for why the ammonia feature in Eps~Ind~Ab is shallower than models predict, and may confirm whether its atmosphere contains thick water-ice clouds.

Increasing the sample of cold exoplanets with photometric and/or spectroscopic data in the upcoming years will provide more insight into the key physics of these cold atmospheres. In particular, it would be highly valuable to measure F1065C and F1140C photometry of other extremely cold exoplanets and brown dwarfs, and determine whether these have similarly shallow ammonia features as Eps~Ind~Ab and WISE 0855 and when the ammonia feature starts to diverge from model predictions. Since the discovery of Eps~Ind~Ab, two other cold exoplanets have been imaged with JWST: 14~Her~c \citep{BardalezGagliuffi2025} and TWA~7~b \citep{Lagrange2025, Crotts2025}; MIRI multi-band photometry of both objects would be highly valuable in confirming the depth of their ammonia features. Strikingly, both of these exoplanets have fainter-than-expected 4.4\micron~magnitudes, based on NIRCam coronagraphic observations; Eps~Ind~Ab has no published NIRCam observations but NaCo non-detections confirm that this planet is also fainter than expected in the 3-5\micron~wavelength range. Together, this sample of cold planets is painting a consistent picture: cold planets are very faint between 3-5\micron, which may be due to the presence of water-ice clouds, and this is an important consideration when designing future photometric searches for cold exoplanets.

\begin{acknowledgments}

We thank Ewan Douglas, Dominique Petit dit de la Roche, and Florian Philipot for their contributions to the proposal on which this work is based.
We also thank the anonymous referee for a thorough report which has certainly improved the quality of the paper.

This work is based on observations made with the NASA/ESA/CSA James Webb Space Telescope. The data were obtained from the Mikulski Archive for Space Telescopes at the Space Telescope Science Institute, which is operated by the Association of Universities for Research in Astronomy, Inc., under NASA contract NAS 5-03127 for JWST. These observations are associated with program \#5037 and \#2243, of which support was provided by NASA through a grant from the Space Telescope Science Institute, which is operated by the Association of Universities for Research in Astronomy, Inc., under NASA contract NAS 5-03127.

J.M. acknowledges support from the National Science Foundation Graduate Research Fellowship Program under Grant No. DGE 2137420.
L.A.B. acknowledges support from the Dutch Research Council (NWO) under grant VI.Veni.242.055 (\url{https://doi.org/10.61686/LAJVP77714}).
J.A.B. acknowledges that part of this research was carried out at the Jet Propulsion Laboratory, California Institute of Technology, under contract with the National Aeronautics and Space Administration (NASA).

\end{acknowledgments}

\begin{contribution}

ECM designed the project and led the observational design, data analysis and interpretation for this paper, and wrote the manuscript. ECM also led the observing proposal upon which this paper is based, with support from LB, JAB, ALC, IJMC, ED, FF, AML, CVM, FP and MWP. JM provided cold model atmospheres and color-magnitude diagram data, as well as contributions to the interpretation and text for Section \ref{sec:ammonia}, with support from CM. MM and ALC contributed to the data reduction. BR contributed to the orbit fitting analysis. Finally, all authors read the manuscript and provided comments.

\end{contribution}

\facilities{JWST (MIRI)}

\software{\texttt{spaceKLIP} \citep{Kammerer2022,Carter2023},
        \texttt{STPSF} \citep{Perrin2014},
        \texttt{PICASO} \citep{Batalha2019, Mukherjee2023,Mang2026},
        \texttt{Virga} \citep{virga_code, virga_paper1, virga_paper2},
        \texttt{astropy} \citep{astropy_i,astropy_ii,astropy_iii},
        \texttt{scipy} \citep{scipy},
        \texttt{numpy} \citep{numpy},
        \texttt{pandas} \citep{pandas_i,pandas_ii},
        \texttt{matplotlib} \citep{matplotlib}, 
        \texttt{seaborn} \citep{seaborn}.
}

\appendix

\section{Archival brown dwarf \& planet photometry}
\label{sec:archivalbds}

In this Section we provide the references and inclusion criteria for the brown dwarfs and planets shown in Figures \ref{fig:ammonia_cmd}-\ref{fig:ammonia_cmd_depletion}. No brown dwarfs have been observed photometrically with the MIRI coronagraphic filters, but we used spectroscopic observations to generate synthetic photometry of several field brown dwarfs for our CMD. For this we used spectroscopic observations of two main samples of cool brown dwarfs: the archive of ultracool brown dwarfs observed with Spitzer/IRS as assembled in \citet{Suarez2022}, and sample of T- and Y-type brown dwarfs observed with JWST/LRS by \citet{Beiler2024b}. In order to improve the visual clarity of Figure \ref{fig:ammonia_cmd}, we restricted the \citet{Suarez2022} sample to targets with a distance precision of 5\% or better, while for the \citet{Beiler2024b} sample we included all targets. We calculated photometry and associated uncertainties by integrating these spectra through the relevant JWST filters and through the NaCo L' filter, making the optimistic assumption that uncertainties on individual bins are uncorrelated, and combined all uncertainties in quadrature. Note that not all of these brown dwarfs have spectral coverage across the full relevant wavelength range, and these appear in only a subset of panels in Figure \ref{fig:ammonia_cmd}.

There are also a handful of MIRI/MRS observations of cold brown dwarfs, which provide the best combination of sensitivity and wavelength coverage for any cold substellar objects, and in particular can be used to calculate synthetic photometry for cold brown dwarfs in the 15.5\micron~MIRI coronagraphic filter. The relevant targets are WISE 0855 \citep{Kuhnle2025}, WISE 1828 \citep{Barrado2023}, WISE 1738 \citep{Vasist2025}, WISE 0458 \citep{Matthews2025}, Ross 458C (Whiteford et al.~in prep.), VHS 1256 b \citep{Miles2023}, PSO J318 \citep{Molliere2025}, 2MASS2148 and 2MASS0624 (priv. comm., from program GO \#2288, PI Lothringer). In each case, we used the data reductions from the published papers, since there are systematic biases in the MAST data reductions (see Appendix \ref{sec:datareductionbias} for details). Where possible, we supplemented these observations with near-IR JWST spectra so as to derive a NaCo L' magnitude. These are available in the same publications as the MRS data for Ross 458C, VHS 1256 b, PSO J318, 2MASS2148 and 2MASS0624, and in dedicated publications for WISE 0855 \citep{Luhman2024} and WISE 1828 \citep{Lew2024}.

Figures \ref{fig:ammonia_cmd}-\ref{fig:ammonia_cmd_depletion} also highlight the CMD positions of bound substellar companions for which both F1065C and F1140C coronagraphic observations are available. We focus only on systems with published observations in both the F1065C and F1140C MIRI filters, since our primary focus in this analysis is the ammonia feature. There are five other systems (with eight companions) that meet this criterion: HR8799bcde \citep{Boccaletti2023}, HD95086b \citep{Malin2024}, GJ504b \citep{Malin2025}, HR2562B \citep{Godoy2024} and Kap And b \citep{Godoy2025}. We supplemented these data with L' photometry collected from the ground for targets where these are available, namely HR8799bcde \citep{Marois2008,Marois2010}, GJ504b \citep{Kuzuhara2013}, HD95086b \citep{Rameau2013} and Kap And b \citep{Bonnefoy2014}.

\section{\ce{H2O} and \ce{NH3} condensation}
\label{sec:condensation}

\begin{figure}
    \centering
    \includegraphics[width=0.49\textwidth]{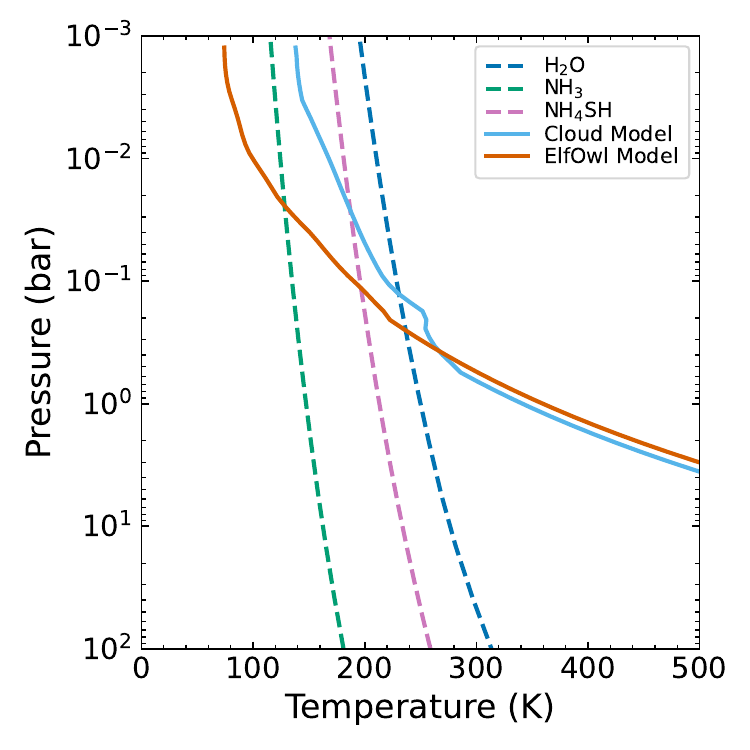}
    \caption{Condensation curves for \ce{H2O}, \ce{NH4SH} and \ce{NH3}, plotted against two pressure-temperature profiles: (1) a 275\,K Sonora ElfOwl model with elevated metallicity ([Fe/H]=1.0) and C/O ($2.5\times$ solar), for which all three cloud species could form, and (2) our custom cloud model from Figure \ref{fig:cloud_model}, for which \ce{H2O} and \ce{NH4SH} but \textit{not} \ce{NH3} clouds are expected to form. The \ce{NH4SH} curve is plotted following the equation provided in \citet{Visscher2006}, for metallicity $3\times$ solar.}
    \label{fig:condensation}
\end{figure}

In Figure \ref{fig:condensation} we provide condensation curves for both \ce{H2O} and \ce{NH3}, alongside the pressure-temperature (P-T) profile for several atmosphere models. First, we show the custom cloud model from Figure \ref{fig:cloud_model}; in this scenario \ce{H2O} condenses but \ce{NH3} condensation is not expected. Second, we show the best-fit model presented in \citet{Matthews2024}: a model from the Sonora Elf Owl suite \citep{Mukherjee2024}, with temperature 275\,K, $10\times$ solar metallicity, $2.5\times$ solar C/O, and $K_{\rm zz}=10^9$\,cm$^2$/s. In this scenario \ce{H2O} condenses at $\sim0.3$\,bar, and \ce{NH3} condenses at $\sim0.02$\,bar, and so both cloud species should be able to form. The production of ammonia clouds in Eps Ind Ab depends finely on the exact atmospheric state, and future works should explore this question in more detail, both observationally and theoretically.

\section{Biases in MIRI/MRS pre-reduced MAST spectra}
\label{sec:datareductionbias}

\begin{figure}
    \centering
    \includegraphics[width=0.49\textwidth]{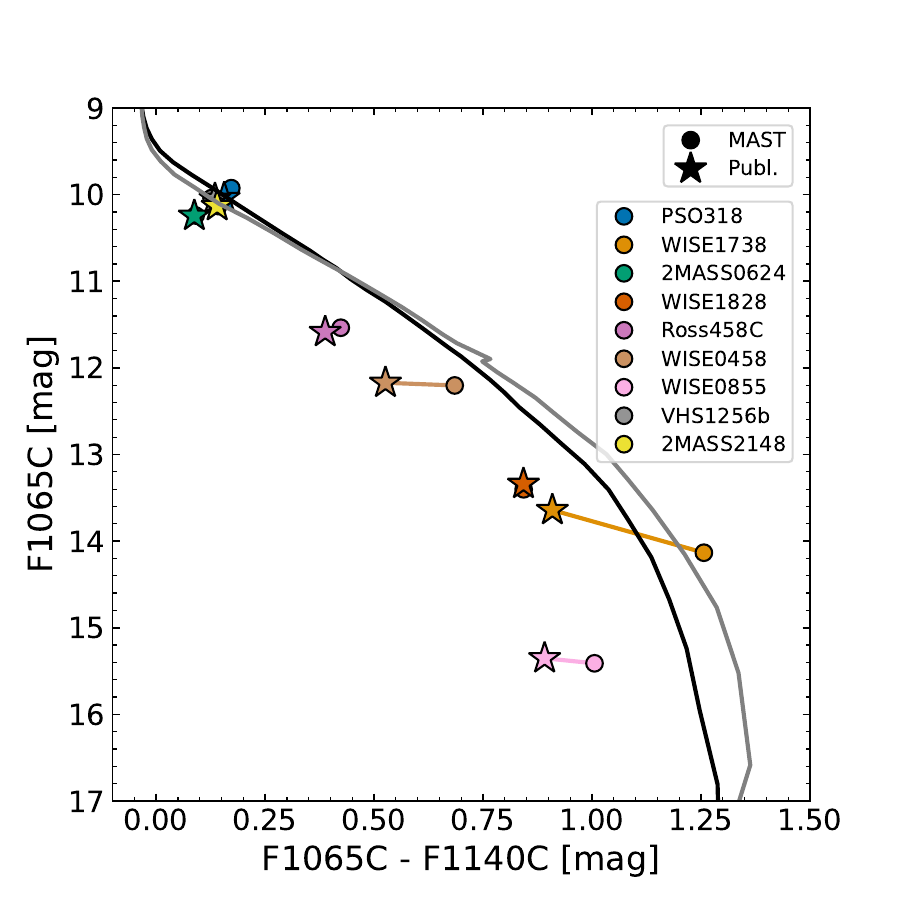}
    \caption{Comparison between CMD positions based on default MAST spectra (circles) and custom data reductions (stars) which are expected to be more reliable for these faint brown dwarfs. This demonstrates the importance of carefully considering systematics when reducing JWST/MRS data. For reference, Sonora Flame Skimmer models with [Fe/H]=0.0 (black) and 0.5 (grey) are indicated: the biases in MAST spectra are sufficient that they can significantly impact which model is deemed the best fit to JWST data.}
    \label{fig:mast_vs_publ}
\end{figure}

We used MIRI/MRS archival spectra of cold brown dwarfs as part of our effort to construct observational CMDs. While compiling this data, we found a significant difference between the CMD positions of pre-reduced spectra acquired directly from MAST and from peer-reviewed papers presenting each individual spectrum. This is demonstrated in Figure \ref{fig:mast_vs_publ}, where we show an F1065C-F1140C color based on MIRI/MRS observations, as derived from either the MAST data reductions or the published data reductions. The effect is most pronounced for the coldest brown dwarf observations (e.g. WISE0458, WISE1828 and especially WISE1738), likely because these are the faintest brown dwarfs and any systematic effects can therefore bias the spectrum most strongly. Further, we see that MAST data reductions consistently imply a larger F1140C-F1065C color than the published data reductions. 

Presumably, this is due to custom data reduction choices made in the more detailed analyses for published papers, such as whether the background is subtracted based on an annulus in the same observation or based on a dithered observation of the target \citep[see e.g.][]{Barrado2023,Kuhnle2025}, and any custom routines that are used to reduce/remove systematics \citep[e.g.][]{Miles2023,Matthews2025}. By nature, the default MAST reductions are designed to be applicable to as broad a range of observations as possible, and are not customized to specific science cases.
This divergence between the MAST spectra and the published spectra serves as a reminder of the need for dedicated data reductions, especially for low signal-to-noise targets, and careful checks to ensure that systematics in the data are understood and removed where possible. \\

\section{Correlations of orbit fitting parameters}
\label{sec:orbit_corner}

\begin{figure*}
    \centering
    \includegraphics[width=0.95\textwidth]{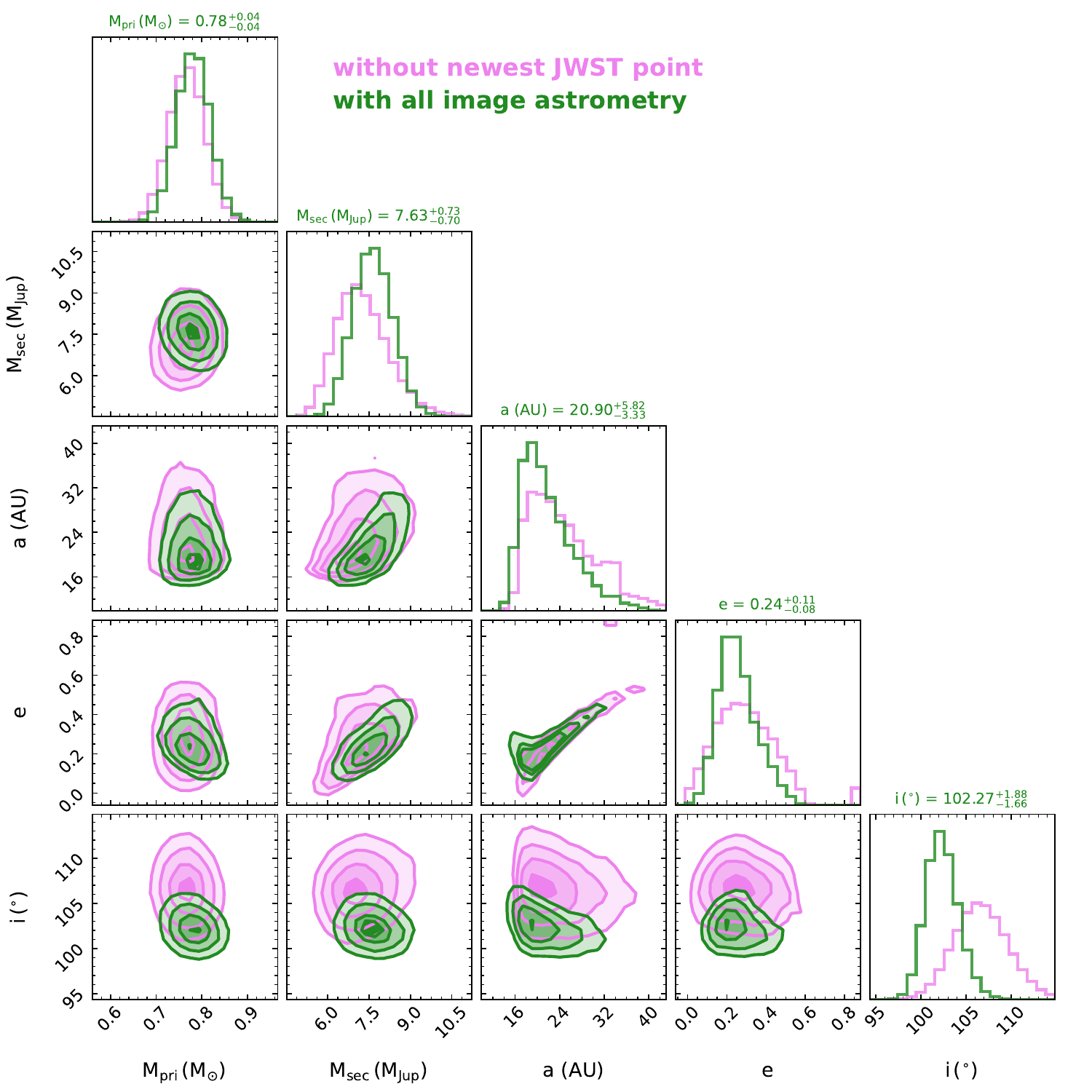}
    \caption{Posterior correlations for key orbit parameters. Green (pink) posteriors show values with (without) the latest JWST point; the orbit fits are otherwise identical. Parameters above each histogram correspond to the best-fit values with the newest JWST image astrometry included.}
    \label{fig:orbit_corner}
\end{figure*}

In Figure \ref{fig:orbit_corner} we include a corner plot showing the correlations between key orbit parameters, as derived using \texttt{orvara} \citep{BrandtOrvara}. We additionally show posteriors without the newest imaging point, to demonstrate how this imaging epoch impacts the parameter values. 

\bibliography{elisabeth.bib}{}

\begin{thebibliography}{}
\expandafter\ifx\csname natexlab\endcsname\relax\def\natexlab#1{#1}\fi
\providecommand{\url}[1]{\href{#1}{#1}}
\providecommand{\dodoi}[1]{doi:~\href{http://doi.org/#1}{\nolinkurl{#1}}}
\providecommand{\doeprint}[1]{\href{http://ascl.net/#1}{\nolinkurl{http://ascl.net/#1}}}
\providecommand{\doarXiv}[1]{\href{https://arxiv.org/abs/#1}{\nolinkurl{https://arxiv.org/abs/#1}}}

\bibitem[{A.~S. {Ackerman} \& M.~S. {Marley}(2001){Ackerman} \&
  {Marley}}]{Ackerman2001}
{Ackerman}, A.~S., \& {Marley}, M.~S. 2001, \bibinfo{title}{{Precipitating
  Condensation Clouds in Substellar Atmospheres},} \apj, 556, 872,
  \dodoi{10.1086/321540}

\bibitem[{ {Astropy Collaboration} {et~al.}(2013){Astropy Collaboration},
  {Robitaille}, {Tollerud}, {Greenfield}, {Droettboom}, {Bray}, {Aldcroft},
  {Davis}, {Ginsburg}, {Price-Whelan}, {Kerzendorf}, {Conley}, {Crighton},
  {Barbary}, {Muna}, {Ferguson}, {Grollier}, {Parikh}, {Nair}, {Unther},
  {Deil}, {Woillez}, {Conseil}, {Kramer}, {Turner}, {Singer}, {Fox}, {Weaver},
  {Zabalza}, {Edwards}, {Azalee Bostroem}, {Burke}, {Casey}, {Crawford},
  {Dencheva}, {Ely}, {Jenness}, {Labrie}, {Lim}, {Pierfederici}, {Pontzen},
  {Ptak}, {Refsdal}, {Servillat}, \& {Streicher}}]{astropy_i}
{Astropy Collaboration}, {Robitaille}, T.~P., {Tollerud}, E.~J., {et~al.} 2013,
  \bibinfo{title}{{Astropy: A community Python package for astronomy},} \aap,
  558, A33, \dodoi{10.1051/0004-6361/201322068}

\bibitem[{ {Astropy Collaboration} {et~al.}(2018){Astropy Collaboration},
  {Price-Whelan}, {Sip{\H{o}}cz}, {G{\"u}nther}, {Lim}, {Crawford}, {Conseil},
  {Shupe}, {Craig}, {Dencheva}, {Ginsburg}, {VanderPlas}, {Bradley},
  {P{\'e}rez-Su{\'a}rez}, {de Val-Borro}, {Aldcroft}, {Cruz}, {Robitaille},
  {Tollerud}, {Ardelean}, {Babej}, {Bach}, {Bachetti}, {Bakanov}, {Bamford},
  {Barentsen}, {Barmby}, {Baumbach}, {Berry}, {Biscani}, {Boquien}, {Bostroem},
  {Bouma}, {Brammer}, {Bray}, {Breytenbach}, {Buddelmeijer}, {Burke},
  {Calderone}, {Cano Rodr{\'\i}guez}, {Cara}, {Cardoso}, {Cheedella}, {Copin},
  {Corrales}, {Crichton}, {D'Avella}, {Deil}, {Depagne}, {Dietrich}, {Donath},
  {Droettboom}, {Earl}, {Erben}, {Fabbro}, {Ferreira}, {Finethy}, {Fox},
  {Garrison}, {Gibbons}, {Goldstein}, {Gommers}, {Greco}, {Greenfield},
  {Groener}, {Grollier}, {Hagen}, {Hirst}, {Homeier}, {Horton}, {Hosseinzadeh},
  {Hu}, {Hunkeler}, {Ivezi{\'c}}, {Jain}, {Jenness}, {Kanarek}, {Kendrew},
  {Kern}, {Kerzendorf}, {Khvalko}, {King}, {Kirkby}, {Kulkarni}, {Kumar},
  {Lee}, {Lenz}, {Littlefair}, {Ma}, {Macleod}, {Mastropietro}, {McCully},
  {Montagnac}, {Morris}, {Mueller}, {Mumford}, {Muna}, {Murphy}, {Nelson},
  {Nguyen}, {Ninan}, {N{\"o}the}, {Ogaz}, {Oh}, {Parejko}, {Parley}, {Pascual},
  {Patil}, {Patil}, {Plunkett}, {Prochaska}, {Rastogi}, {Reddy Janga},
  {Sabater}, {Sakurikar}, {Seifert}, {Sherbert}, {Sherwood-Taylor}, {Shih},
  {Sick}, {Silbiger}, {Singanamalla}, {Singer}, {Sladen}, {Sooley},
  {Sornarajah}, {Streicher}, {Teuben}, {Thomas}, {Tremblay}, {Turner},
  {Terr{\'o}n}, {van Kerkwijk}, {de la Vega}, {Watkins}, {Weaver}, {Whitmore},
  {Woillez}, {Zabalza}, \& {Astropy Contributors}}]{astropy_ii}
{Astropy Collaboration}, {Price-Whelan}, A.~M., {Sip{\H{o}}cz}, B.~M., {et~al.}
  2018, \bibinfo{title}{{The Astropy Project: Building an Open-science Project
  and Status of the v2.0 Core Package},} \aj, 156, 123,
  \dodoi{10.3847/1538-3881/aabc4f}

\bibitem[{ {Astropy Collaboration} {et~al.}(2022){Astropy Collaboration},
  {Price-Whelan}, {Lim}, {Earl}, {Starkman}, {Bradley}, {Shupe}, {Patil},
  {Corrales}, {Brasseur}, {N{\"o}the}, {Donath}, {Tollerud}, {Morris},
  {Ginsburg}, {Vaher}, {Weaver}, {Tocknell}, {Jamieson}, {van Kerkwijk},
  {Robitaille}, {Merry}, {Bachetti}, {G{\"u}nther}, {Aldcroft},
  {Alvarado-Montes}, {Archibald}, {B{\'o}di}, {Bapat}, {Barentsen},
  {Baz{\'a}n}, {Biswas}, {Boquien}, {Burke}, {Cara}, {Cara}, {Conroy},
  {Conseil}, {Craig}, {Cross}, {Cruz}, {D'Eugenio}, {Dencheva}, {Devillepoix},
  {Dietrich}, {Eigenbrot}, {Erben}, {Ferreira}, {Foreman-Mackey}, {Fox},
  {Freij}, {Garg}, {Geda}, {Glattly}, {Gondhalekar}, {Gordon}, {Grant},
  {Greenfield}, {Groener}, {Guest}, {Gurovich}, {Handberg}, {Hart},
  {Hatfield-Dodds}, {Homeier}, {Hosseinzadeh}, {Jenness}, {Jones}, {Joseph},
  {Kalmbach}, {Karamehmetoglu}, {Ka{\l}uszy{\'n}ski}, {Kelley}, {Kern},
  {Kerzendorf}, {Koch}, {Kulumani}, {Lee}, {Ly}, {Ma}, {MacBride}, {Maljaars},
  {Muna}, {Murphy}, {Norman}, {O'Steen}, {Oman}, {Pacifici}, {Pascual},
  {Pascual-Granado}, {Patil}, {Perren}, {Pickering}, {Rastogi}, {Roulston},
  {Ryan}, {Rykoff}, {Sabater}, {Sakurikar}, {Salgado}, {Sanghi}, {Saunders},
  {Savchenko}, {Schwardt}, {Seifert-Eckert}, {Shih}, {Jain}, {Shukla}, {Sick},
  {Simpson}, {Singanamalla}, {Singer}, {Singhal}, {Sinha}, {Sip{\H{o}}cz},
  {Spitler}, {Stansby}, {Streicher}, {{\v{S}}umak}, {Swinbank}, {Taranu},
  {Tewary}, {Tremblay}, {de Val-Borro}, {Van Kooten}, {Vasovi{\'c}}, {Verma},
  {de Miranda Cardoso}, {Williams}, {Wilson}, {Winkel}, {Wood-Vasey}, {Xue},
  {Yoachim}, {Zhang}, {Zonca}, \& {Astropy Project Contributors}}]{astropy_iii}
{Astropy Collaboration}, {Price-Whelan}, A.~M., {Lim}, P.~L., {et~al.} 2022,
  \bibinfo{title}{{The Astropy Project: Sustaining and Growing a
  Community-oriented Open-source Project and the Latest Major Release (v5.0) of
  the Core Package},} \apj, 935, 167, \dodoi{10.3847/1538-4357/ac7c74}

\bibitem[{D.~C. {Bardalez Gagliuffi} {et~al.}(2025){Bardalez Gagliuffi},
  {Balmer}, {Pueyo}, {Brandt}, {Giovinazzi}, {Millholland}, {Black}, {Lu},
  {Rice}, {Mang}, {Morley}, {Lacy}, {Girard}, {Matthews}, {Carter}, {Bowler},
  {Faherty}, {Fontanive}, \& {Rickman}}]{BardalezGagliuffi2025}
{Bardalez Gagliuffi}, D.~C., {Balmer}, W.~O., {Pueyo}, L., {et~al.} 2025,
  \bibinfo{title}{{JWST Coronagraphic Images of 14 Her c: A Cold Giant Planet
  in a Dynamically Hot Multiplanet System},} \apjl, 988, L18,
  \dodoi{10.3847/2041-8213/ade30f}

\bibitem[{D. {Barrado} {et~al.}(2023){Barrado}, {Molli{\`e}re}, {Patapis},
  {Min}, {Tremblin}, {Ardevol Martinez}, {Whiteford}, {Vasist}, {Argyriou},
  {Samland}, {Lagage}, {Decin}, {Waters}, {Henning}, {Morales-Calder{\'o}n},
  {Guedel}, {Vandenbussche}, {Absil}, {Baudoz}, {Boccaletti}, {Bouwman},
  {Cossou}, {Coulais}, {Crouzet}, {Gastaud}, {Glasse}, {Glauser}, {Kamp},
  {Kendrew}, {Krause}, {Lahuis}, {Mueller}, {Olofsson}, {Pye}, {Rouan},
  {Royer}, {Scheithauer}, {Waldmann}, {Colina}, {van Dishoeck}, {Ray},
  {{\"O}stlin}, \& {Wright}}]{Barrado2023}
{Barrado}, D., {Molli{\`e}re}, P., {Patapis}, P., {et~al.} 2023,
  \bibinfo{title}{{$^{15}$NH$_{3}$ in the atmosphere of a cool brown dwarf},}
  \nat, 624, 263, \dodoi{10.1038/s41586-023-06813-y}

\bibitem[{N. Batalha {et~al.}(2020)Batalha, caoimherooney11, \&
  sagnickm}]{virga_code}
Batalha, N., caoimherooney11, \& sagnickm. 2020, natashabatalha/virga: Initial
  Release, v0.0 Zenodo, \dodoi{10.5281/zenodo.3759888}

\bibitem[{N.~E. {Batalha} {et~al.}(2019){Batalha}, {Marley}, {Lewis}, \&
  {Fortney}}]{Batalha2019}
{Batalha}, N.~E., {Marley}, M.~S., {Lewis}, N.~K., \& {Fortney}, J.~J. 2019,
  \bibinfo{title}{{Exoplanet Reflected-light Spectroscopy with PICASO},} \apj,
  878, 70, \dodoi{10.3847/1538-4357/ab1b51}

\bibitem[{N.~E. {Batalha} {et~al.}(2025){Batalha}, {Rooney}, {Visscher},
  {Moran}, {Marley}, {Sengupta}, {Kiefer}, {Lodge}, {Mang}, {Morley},
  {Mukherjee}, {Fortney}, {Gao}, {Lewis}, {Mayorga}, {Pearce}, \&
  {Wakeford}}]{virga_paper1}
{Batalha}, N.~E., {Rooney}, C.~M., {Visscher}, C., {et~al.} 2025,
  \bibinfo{title}{{Condensation Clouds in Substellar Atmospheres with Virga},}
  arXiv e-prints, arXiv:2508.15102, \dodoi{10.48550/arXiv.2508.15102}

\bibitem[{S.~A. {Beiler} {et~al.}(2024){Beiler}, {Mukherjee}, {Cushing},
  {Kirkpatrick}, {Schneider}, {Kothari}, {Marley}, \& {Visscher}}]{Beiler2024b}
{Beiler}, S.~A., {Mukherjee}, S., {Cushing}, M.~C., {et~al.} 2024,
  \bibinfo{title}{{A Tale of Two Molecules: The Underprediction of CO$_{2}$ and
  Overprediction of PH$_{3}$ in Late T and Y Dwarf Atmospheric Models},} \apj,
  973, 60, \dodoi{10.3847/1538-4357/ad6759}

\bibitem[{O. {Berne} {et~al.}(2025){Berne}, {Amiot}, {Canin}, \&
  {Schroetter}}]{Berne_jwstgo8438}
{Berne}, O., {Amiot}, P., {Canin}, A., \& {Schroetter}, I. 2025, {The first
  spectrum of a temperate super-Jupiter planet},, JWST Proposal. Cycle 4, ID.
  \#8438

\bibitem[{A. {Boccaletti} {et~al.}(2015){Boccaletti}, {Lagage}, {Baudoz},
  {Beichman}, {Bouchet}, {Cavarroc}, {Dubreuil}, {Glasse}, {Glauser}, {Hines},
  {Lajoie}, {Lebreton}, {Perrin}, {Pueyo}, {Reess}, {Rieke}, {Ronayette},
  {Rouan}, {Soummer}, \& {Wright}}]{Boccaletti2015}
{Boccaletti}, A., {Lagage}, P.~O., {Baudoz}, P., {et~al.} 2015,
  \bibinfo{title}{{The Mid-Infrared Instrument for the James Webb Space
  Telescope, V: Predicted Performance of the MIRI Coronagraphs},} \pasp, 127,
  633, \dodoi{10.1086/682256}

\bibitem[{A. {Boccaletti} {et~al.}(2022){Boccaletti}, {Cossou}, {Baudoz},
  {Lagage}, {Dicken}, {Glasse}, {Hines}, {Aguilar}, {Detre}, {Nickson},
  {Noriega-Crespo}, {G{\'a}sp{\'a}r}, {Labiano}, {Stark}, {Rouan}, {Reess},
  {Wright}, {Rieke}, {Garcia Marin}, {Lajoie}, {Girard}, {Perrin}, {Soummer},
  \& {Pueyo}}]{Boccaletti2022}
{Boccaletti}, A., {Cossou}, C., {Baudoz}, P., {et~al.} 2022,
  \bibinfo{title}{{JWST/MIRI coronagraphic performances as measured on-sky},}
  \aap, 667, A165, \dodoi{10.1051/0004-6361/202244578}

\bibitem[{A. {Boccaletti} {et~al.}(2023){Boccaletti}, {M{\^a}lin}, {Baudoz},
  {Tremplin}, {Perrot}, {Rouan}, {-O.}, {Whiteford}, {Molli{\`e}re}, {Waters},
  {Henning}, {Decin}, {G{\"u}del}, {Vadenbussche}, {Absil}, {Argyriou},
  {Bouwman}, {Cossou}, {Coulais}, {Gastaud}, {Glasse}, {Glauser}, {Kamp},
  {Kendrew}, {Krause}, {Lahuis}, {Mueller}, {Olofsson}, {Patapis}, {Pye},
  {Royer}, {Serabyn}, {Scheithauer}, {Colina}, {van Dischoeck E.}, {Ostlin},
  {T.}, \& {G}}]{Boccaletti2023}
{Boccaletti}, A., {M{\^a}lin}, M., {Baudoz}, P., {et~al.} 2023,
  \bibinfo{title}{{Imaging detection of the inner dust belt and the four
  exoplanets in the HR8799 system with JWST's MIRI coronagraph},} arXiv
  e-prints, arXiv:2310.13414, \dodoi{10.48550/arXiv.2310.13414}

\bibitem[{M. {Bonnefoy} {et~al.}(2014){Bonnefoy}, {Currie}, {Marleau},
  {Schlieder}, {Wisniewski}, {Carson}, {Covey}, {Henning}, {Biller}, {Hinz},
  {Klahr}, {Marsh Boyer}, {Zimmerman}, {Janson}, {McElwain}, {Mordasini},
  {Skemer}, {Bailey}, {Defr{\`e}re}, {Thalmann}, {Skrutskie}, {Allard},
  {Homeier}, {Tamura}, {Feldt}, {Cumming}, {Grady}, {Brandner}, {Helling},
  {Witte}, {Hauschildt}, {Kandori}, {Kuzuhara}, {Fukagawa}, {Kwon}, {Kudo},
  {Hashimoto}, {Kusakabe}, {Abe}, {Brandt}, {Egner}, {Guyon}, {Hayano},
  {Hayashi}, {Hayashi}, {Hodapp}, {Ishii}, {Iye}, {Knapp}, {Matsuo}, {Mede},
  {Miyama}, {Morino}, {Moro-Martin}, {Nishimura}, {Pyo}, {Serabyn}, {Suenaga},
  {Suto}, {Suzuki}, {Takahashi}, {Takami}, {Takato}, {Terada}, {Tomono},
  {Turner}, {Watanabe}, {Yamada}, {Takami}, \& {Usuda}}]{Bonnefoy2014}
{Bonnefoy}, M., {Currie}, T., {Marleau}, G.-D., {et~al.} 2014,
  \bibinfo{title}{{Characterization of the gaseous companion
  {\ensuremath{\kappa}} Andromedae b. New Keck and LBTI high-contrast
  observations},} \aap, 562, A111, \dodoi{10.1051/0004-6361/201322119}

\bibitem[{B.~P. {Bowler}(2016){Bowler}}]{Bowler2016}
{Bowler}, B.~P. 2016, \bibinfo{title}{{Imaging Extrasolar Giant Planets},}
  \pasp, 128, 102001, \dodoi{10.1088/1538-3873/128/968/102001}

\bibitem[{T.~D. {Brandt}(2018){Brandt}}]{Brandt2018}
{Brandt}, T.~D. 2018, \bibinfo{title}{{The Hipparcos-Gaia Catalog of
  Accelerations},} \apjs, 239, 31, \dodoi{10.3847/1538-4365/aaec06}

\bibitem[{T.~D. {Brandt}(2021){Brandt}}]{Brandt2021}
{Brandt}, T.~D. 2021, \bibinfo{title}{{The Hipparcos-Gaia Catalog of
  Accelerations: Gaia EDR3 Edition},} \apjs, 254, 42,
  \dodoi{10.3847/1538-4365/abf93c}

\bibitem[{T.~D. {Brandt} {et~al.}(2021){Brandt}, {Dupuy}, {Li}, {Brandt},
  {Zeng}, {Michalik}, {Bardalez Gagliuffi}, \& {Raposo-Pulido}}]{BrandtOrvara}
{Brandt}, T.~D., {Dupuy}, T.~J., {Li}, Y., {et~al.} 2021,
  \bibinfo{title}{{orvara: An Efficient Code to Fit Orbits Using Radial
  Velocity, Absolute, and/or Relative Astrometry},} \aj, 162, 186,
  \dodoi{10.3847/1538-3881/ac042e}

\bibitem[{H. Bushouse {et~al.}(2023)Bushouse, Eisenhamer, Dencheva, Davies,
  Greenfield, Morrison, Hodge, Simon, Grumm, Droettboom, Slavich, Sosey, Pauly,
  Miller, Jedrzejewski, Hack, Davis, Crawford, Law, Gordon, Regan, Cara,
  MacDonald, Bradley, Shanahan, Jamieson, Teodoro, Williams, \&
  Pena-Guerrero}]{Bushouse_jwst}
Bushouse, H., Eisenhamer, J., Dencheva, N., {et~al.} 2023, JWST Calibration
  Pipeline, 1.12.5 Zenodo, \dodoi{10.5281/zenodo.10022973}

\bibitem[{A.~L. {Carter} {et~al.}(2023){Carter}, {Hinkley}, {Kammerer},
  {Skemer}, {Biller}, {Leisenring}, {Millar-Blanchaer}, {Petrus}, {Stone},
  {Ward-Duong}, {Wang}, {Girard}, {Hines}, {Perrin}, {Pueyo}, {Balmer},
  {Bonavita}, {Bonnefoy}, {Chauvin}, {Choquet}, {Christiaens}, {Danielski},
  {Kennedy}, {Matthews}, {Miles}, {Patapis}, {Ray}, {Rickman}, {Sallum},
  {Stapelfeldt}, {Whiteford}, {Zhou}, {Absil}, {Boccaletti}, {Booth}, {Bowler},
  {Chen}, {Currie}, {Fortney}, {Grady}, {Greebaum}, {Henning}, {Hoch},
  {Janson}, {Kalas}, {Kenworthy}, {Kervella}, {Kraus}, {Lagage}, {Liu},
  {Macintosh}, {Marino}, {Marley}, {Marois}, {Matthews}, {Mawet}, {McElwain},
  {Metchev}, {Meyer}, {Molliere}, {Moran}, {Morley}, {Mukherjee}, {Pantin},
  {Quirrenbach}, {Rebollido}, {Ren}, {Schneider}, {Vasist}, {Worthen}, {Wyatt},
  {Briesemeister}, {Bryan}, {Calissendorff}, {Cantalloube}, {Cugno}, {De
  Furio}, {Dupuy}, {Factor}, {Faherty}, {Fitzgerald}, {Franson}, {Gonzales},
  {Hood}, {Howe}, {Kuzuhara}, {Lagrange}, {Lawson}, {Lazzoni}, {Lew}, {Liu},
  {Llop-Sayson}, {Lloyd}, {Martinez}, {Mazoyer}, {Palma-Bifani}, {Quanz},
  {Redai}, {Samland}, {Schlieder}, {Tamura}, {Tan}, {Uyama}, {Vigan}, {Vos},
  {Wagner}, {Wolff}, {Ygouf}, {Zhang}, {Zhang}, \& {Zhang}}]{Carter2023}
{Carter}, A.~L., {Hinkley}, S., {Kammerer}, J., {et~al.} 2023,
  \bibinfo{title}{{The JWST Early Release Science Program for Direct
  Observations of Exoplanetary Systems I: High-contrast Imaging of the
  Exoplanet HIP 65426 b from 2 to 16 {\ensuremath{\mu}}m},} \apjl, 951, L20,
  \dodoi{10.3847/2041-8213/acd93e}

\bibitem[{P.~A. {Coles} {et~al.}(2019){Coles}, {Yurchenko}, \&
  {Tennyson}}]{Coles2019}
{Coles}, P.~A., {Yurchenko}, S.~N., \& {Tennyson}, J. 2019,
  \bibinfo{title}{{ExoMol molecular line lists - XXXV. A rotation-vibration
  line list for hot ammonia},} \mnras, 490, 4638, \dodoi{10.1093/mnras/stz2778}

\bibitem[{K.~A. {Crotts} {et~al.}(2025){Crotts}, {Carter}, {Lawson}, {Mang},
  {Biller}, {Booth}, {Ferrer-Chavez}, {Girard}, {Lagrange}, {Liu}, {Marino},
  {Millar-Blanchaer}, {Skemer}, {Strampelli}, {Wang}, {Absil}, {Balmer},
  {Bendahan-West}, {Bogat}, {Bowens-Rubin}, {Chauvin}, {Fontanive}, {Franson},
  {Kammerer}, {Leisenring}, {Morley}, {Rebollido}, {Skaf}, {Sutlieff},
  {Bruinsma}, {Hinkley}, {Hoch}, {James}, {Kane}, {Mawet}, {Meyer},
  {Palatnick}, {Perrin}, {Ray}, {Rickman}, {Sanghi}, \&
  {Stephenson}}]{Crotts2025}
{Crotts}, K.~A., {Carter}, A.~L., {Lawson}, K., {et~al.} 2025,
  \bibinfo{title}{{Follow-up Exploration of the TWA 7 Planet{\textendash}Disk
  System with JWST NIRCam},} \apjl, 987, L41, \dodoi{10.3847/2041-8213/ade798}

\bibitem[{M.~C. {Cushing} {et~al.}(2011){Cushing}, {Kirkpatrick}, {Gelino},
  {Griffith}, {Skrutskie}, {Mainzer}, {Marsh}, {Beichman}, {Burgasser},
  {Prato}, {Simcoe}, {Marley}, {Saumon}, {Freedman}, {Eisenhardt}, \&
  {Wright}}]{Cushing2011}
{Cushing}, M.~C., {Kirkpatrick}, J.~D., {Gelino}, C.~R., {et~al.} 2011,
  \bibinfo{title}{{The Discovery of Y Dwarfs using Data from the Wide-field
  Infrared Survey Explorer (WISE)},} \apj, 743, 50,
  \dodoi{10.1088/0004-637X/743/1/50}

\bibitem[{B.-O. {Demory} {et~al.}(2009){Demory}, {S{\'e}gransan}, {Forveille},
  {Queloz}, {Beuzit}, {Delfosse}, {di Folco}, {Kervella}, {Le Bouquin},
  {Perrier}, {Benisty}, {Duvert}, {Hofmann}, {Lopez}, \& {Petrov}}]{Demory2009}
{Demory}, B.-O., {S{\'e}gransan}, D., {Forveille}, T., {et~al.} 2009,
  \bibinfo{title}{{Mass-radius relation of low and very low-mass stars
  revisited with the VLTI},} \aap, 505, 205,
  \dodoi{10.1051/0004-6361/200911976}

\bibitem[{F. {Feng} {et~al.}(2019){Feng}, {Anglada-Escud{\'e}}, {Tuomi},
  {Jones}, {Chanam{\'e}}, {Butler}, \& {Janson}}]{Feng2019}
{Feng}, F., {Anglada-Escud{\'e}}, G., {Tuomi}, M., {et~al.} 2019,
  \bibinfo{title}{{Detection of the nearest Jupiter analogue in radial velocity
  and astrometry data},} \mnras, 490, 5002, \dodoi{10.1093/mnras/stz2912}

\bibitem[{F. {Feng} {et~al.}(2025){Feng}, {Xiao}, {Jones}, {Jenkins}, {Pena},
  \& {Sun}}]{Feng2025}
{Feng}, F., {Xiao}, G.-Y., {Jones}, H. R.~A., {et~al.} 2025,
  \bibinfo{title}{{Lessons learned from the detection of wide companions by
  radial velocity and astrometry},} \mnras, 539, 3180,
  \dodoi{10.1093/mnras/staf689}

\bibitem[{ {Gaia Collaboration} {et~al.}(2021){Gaia Collaboration}, {Brown},
  {Vallenari}, {Prusti}, {de Bruijne}, {Babusiaux}, {Biermann}, {Creevey},
  {Evans}, {Eyer}, {Hutton}, {Jansen}, {Jordi}, {Klioner}, {Lammers},
  {Lindegren}, {Luri}, {Mignard}, {Panem}, {Pourbaix}, {Randich}, {Sartoretti},
  {Soubiran}, {Walton}, {Arenou}, {Bailer-Jones}, {Bastian}, {Cropper},
  {Drimmel}, {Katz}, {Lattanzi}, {van Leeuwen}, {Bakker}, {Cacciari},
  {Casta{\~n}eda}, {De Angeli}, {Ducourant}, {Fabricius}, {Fouesneau},
  {Fr{\'e}mat}, {Guerra}, {Guerrier}, {Guiraud}, {Jean-Antoine Piccolo},
  {Masana}, {Messineo}, {Mowlavi}, {Nicolas}, {Nienartowicz}, {Pailler},
  {Panuzzo}, {Riclet}, {Roux}, {Seabroke}, {Sordo}, {Tanga}, {Th{\'e}venin},
  {Gracia-Abril}, {Portell}, {Teyssier}, {Altmann}, {Andrae}, {Bellas-Velidis},
  {Benson}, {Berthier}, {Blomme}, {Brugaletta}, {Burgess}, {Busso}, {Carry},
  {Cellino}, {Cheek}, {Clementini}, {Damerdji}, {Davidson}, {Delchambre},
  {Dell'Oro}, {Fern{\'a}ndez-Hern{\'a}ndez}, {Galluccio}, {Garc{\'\i}a-Lario},
  {Garcia-Reinaldos}, {Gonz{\'a}lez-N{\'u}{\~n}ez}, {Gosset}, {Haigron},
  {Halbwachs}, {Hambly}, {Harrison}, {Hatzidimitriou}, {Heiter},
  {Hern{\'a}ndez}, {Hestroffer}, {Hodgkin}, {Holl}, {Jan{\ss}en}, {Jevardat de
  Fombelle}, {Jordan}, {Krone-Martins}, {Lanzafame}, {L{\"o}ffler}, {Lorca},
  {Manteiga}, {Marchal}, {Marrese}, {Moitinho}, {Mora}, {Muinonen}, {Osborne},
  {Pancino}, {Pauwels}, {Petit}, {Recio-Blanco}, {Richards}, {Riello},
  {Rimoldini}, {Robin}, {Roegiers}, {Rybizki}, {Sarro}, {Siopis}, {Smith},
  {Sozzetti}, {Ulla}, {Utrilla}, {van Leeuwen}, {van Reeven}, {Abbas}, {Abreu
  Aramburu}, {Accart}, {Aerts}, {Aguado}, {Ajaj}, {Altavilla}, {{\'A}lvarez},
  {{\'A}lvarez Cid-Fuentes}, {Alves}, {Anderson}, {Anglada Varela}, {Antoja},
  {Audard}, {Baines}, {Baker}, {Balaguer-N{\'u}{\~n}ez}, {Balbinot}, {Balog},
  {Barache}, {Barbato}, {Barros}, {Barstow}, {Bartolom{\'e}}, {Bassilana},
  {Bauchet}, {Baudesson-Stella}, {Becciani}, {Bellazzini}, {Bernet}, {Bertone},
  {Bianchi}, {Blanco-Cuaresma}, {Boch}, {Bombrun}, {Bossini}, {Bouquillon},
  {Bragaglia}, {Bramante}, {Breedt}, {Bressan}, {Brouillet}, {Bucciarelli},
  {Burlacu}, {Busonero}, {Butkevich}, {Buzzi}, {Caffau}, {Cancelliere},
  {C{\'a}novas}, {Cantat-Gaudin}, {Carballo}, {Carlucci}, {Carnerero},
  {Carrasco}, {Casamiquela}, {Castellani}, {Castro-Ginard}, {Castro Sampol},
  {Chaoul}, {Charlot}, {Chemin}, {Chiavassa}, {Cioni}, {Comoretto}, {Cooper},
  {Cornez}, {Cowell}, {Crifo}, {Crosta}, {Crowley}, {Dafonte}, {Dapergolas},
  {David}, \& {David}}]{gaia_edr3}
{Gaia Collaboration}, {Brown}, A.~G.~A., {Vallenari}, A., {et~al.} 2021,
  \bibinfo{title}{{Gaia Early Data Release 3. Summary of the contents and
  survey properties},} \aap, 649, A1, \dodoi{10.1051/0004-6361/202039657}

\bibitem[{ {Gaia Collaboration} {et~al.}(2023){Gaia Collaboration},
  {Vallenari}, {Brown}, {Prusti}, {de Bruijne}, {Arenou}, {Babusiaux},
  {Biermann}, {Creevey}, {Ducourant}, \& et~al.}]{gaia_dr3}
{Gaia Collaboration}, {Vallenari}, A., {Brown}, A.~G.~A., {et~al.} 2023,
  \bibinfo{title}{{Gaia Data Release 3. Summary of the content and survey
  properties},} \aap, 674, A1, \dodoi{10.1051/0004-6361/202243940}

\bibitem[{N. {Godoy} {et~al.}(2024){Godoy}, {Choquet}, {Serabyn}, {Danielski},
  {Stolker}, {Charnay}, {Hinkley}, {Lagage}, {Ressler}, {Tremblin}, \&
  {Vigan}}]{Godoy2024}
{Godoy}, N., {Choquet}, E., {Serabyn}, E., {et~al.} 2024, \bibinfo{title}{{A
  new atmospheric characterization of the sub-stellar companion HR 2562 B with
  JWST/MIRI observations},} \aap, 689, A185,
  \dodoi{10.1051/0004-6361/202449951}

\bibitem[{N. {Godoy} {et~al.}(2025){Godoy}, {Choquet}, {Serabyn}, {M{\^a}lin},
  {Tremblin}, {Danielski}, {Lagage}, {Boccaletti}, {Charnay}, \&
  {Ressler}}]{Godoy2025}
{Godoy}, N., {Choquet}, E., {Serabyn}, E., {et~al.} 2025, \bibinfo{title}{{A
  JWST/MIRI view of {\ensuremath{\kappa}} Andromedae b: Refining its mass, age,
  and physical parameters},} \aap, 702, A4, \dodoi{10.1051/0004-6361/202554652}

\bibitem[{J.~J. {Green} {et~al.}(2005){Green}, {Beichman}, {Basinger},
  {Horner}, {Meyer}, {Redding}, {Rieke}, \& {Trauger}}]{Green2005}
{Green}, J.~J., {Beichman}, C., {Basinger}, S.~A., {et~al.} 2005,
  \bibinfo{title}{{High contrast imaging with the JWST NIRCAM coronagraph},} in
  Society of Photo-Optical Instrumentation Engineers (SPIE) Conference Series,
  Vol. 5905, Techniques and Instrumentation for Detection of Exoplanets II, ed.
  D.~R. {Coulter}, 185--195, \dodoi{10.1117/12.619343}

\bibitem[{T. {Guillot} {et~al.}(2020{\natexlab{a}}){Guillot}, {Stevenson},
  {Atreya}, {Bolton}, \& {Becker}}]{Guillot2020a}
{Guillot}, T., {Stevenson}, D.~J., {Atreya}, S.~K., {Bolton}, S.~J., \&
  {Becker}, H.~N. 2020{\natexlab{a}}, \bibinfo{title}{{Storms and the Depletion
  of Ammonia in Jupiter: I. Microphysics of ``Mushballs''},} Journal of
  Geophysical Research (Planets), 125, e06403,
  \dodoi{10.1029/2020JE00640310.1002/essoar.10502154.1}

\bibitem[{T. {Guillot} {et~al.}(2020{\natexlab{b}}){Guillot}, {Li}, {Bolton},
  {Brown}, {Ingersoll}, {Janssen}, {Levin}, {Lunine}, {Orton}, {Steffes}, \&
  {Stevenson}}]{Guillot2020b}
{Guillot}, T., {Li}, C., {Bolton}, S.~J., {et~al.} 2020{\natexlab{b}},
  \bibinfo{title}{{Storms and the Depletion of Ammonia in Jupiter: II.
  Explaining the Juno Observations},} Journal of Geophysical Research
  (Planets), 125, e06404, \dodoi{10.1029/2020JE00640410.1002/essoar.10502179.1}

\bibitem[{C.~R. Harris {et~al.}(2020)Harris, Millman, van~der Walt, Gommers,
  Virtanen, Cournapeau, Wieser, Taylor, Berg, Smith, Kern, Picus, Hoyer, van
  Kerkwijk, Brett, Haldane, del R{\'{i}}o, Wiebe, Peterson,
  G{\'{e}}rard-Marchant, Sheppard, Reddy, Weckesser, Abbasi, Gohlke, \&
  Oliphant}]{numpy}
Harris, C.~R., Millman, K.~J., van~der Walt, S.~J., {et~al.} 2020,
  \bibinfo{title}{Array programming with {NumPy},} Nature, 585, 357,
  \dodoi{10.1038/s41586-020-2649-2}

\bibitem[{J.~D. Hunter(2007)Hunter}]{matplotlib}
Hunter, J.~D. 2007, \bibinfo{title}{Matplotlib: A 2D graphics environment,}
  Computing in Science \& Engineering, 9, 90, \dodoi{10.1109/MCSE.2007.55}

\bibitem[{P.~G.~J. {Irwin} {et~al.}(1998){Irwin}, {Weir}, {Smith}, {Taylor},
  {Lambert}, {Calcutt}, {Cameron-Smith}, {Carlson}, {Baines}, {Orton},
  {Drossart}, {Encrenaz}, \& {Roos-Serote}}]{Irwin1998}
{Irwin}, P.~G.~J., {Weir}, A.~L., {Smith}, S.~E., {et~al.} 1998,
  \bibinfo{title}{{Cloud structure and atmospheric composition of Jupiter
  retrieved from Galileo near-infrared mapping spectrometer real-time
  spectra},} \jgr, 103, 23001, \dodoi{10.1029/98JE00948}

\bibitem[{M. {Janson} {et~al.}(2009){Janson}, {Apai}, {Zechmeister},
  {Brandner}, {K{\"u}rster}, {Kasper}, {Reffert}, {Endl}, {Lafreni{\`e}re},
  {Gei{\ss}ler}, {Hippler}, \& {Henning}}]{Janson2009}
{Janson}, M., {Apai}, D., {Zechmeister}, M., {et~al.} 2009,
  \bibinfo{title}{{Imaging search for the unseen companion to eps Ind A -
  improving the detection limits with 4 {\ensuremath{\mu}}m observations},}
  \mnras, 399, 377, \dodoi{10.1111/j.1365-2966.2009.15285.x}

\bibitem[{J. {Kammerer} {et~al.}(2022){Kammerer}, {Stark}, {Ludwick},
  {Juanola-Parramon}, \& {Nemati}}]{Kammerer2022}
{Kammerer}, J., {Stark}, C.~C., {Ludwick}, K.~J., {Juanola-Parramon}, R., \&
  {Nemati}, B. 2022, \bibinfo{title}{{Simulating Reflected Light Coronagraphy
  of Earth-like Exoplanets with a Large IR/O/UV Space Telescope: Impact and
  Calibration of Smooth Exozodiacal Dust},} \aj, 164, 235,
  \dodoi{10.3847/1538-3881/ac97eb}

\bibitem[{P. {Kervella} {et~al.}(2019){Kervella}, {Arenou}, {Mignard}, \&
  {Th{\'e}venin}}]{Kervella2019}
{Kervella}, P., {Arenou}, F., {Mignard}, F., \& {Th{\'e}venin}, F. 2019,
  \bibinfo{title}{{Stellar and substellar companions of nearby stars from Gaia
  DR2. Binarity from proper motion anomaly},} \aap, 623, A72,
  \dodoi{10.1051/0004-6361/201834371}

\bibitem[{J.~E. {Krist} {et~al.}(2009){Krist}, {Balasubramanian}, {Beichman},
  {Echternach}, {Green}, {Liewer}, {Muller}, {Serabyn}, {Shaklan}, {Trauger},
  {Wilson}, {Horner}, {Mao}, {Somerstein}, {Vasudevan}, {Kelly}, \&
  {Rieke}}]{Krist2009}
{Krist}, J.~E., {Balasubramanian}, K., {Beichman}, C.~A., {et~al.} 2009,
  \bibinfo{title}{{The JWST/NIRCam coronagraph: mask design and fabrication},}
  in Society of Photo-Optical Instrumentation Engineers (SPIE) Conference
  Series, Vol. 7440, Techniques and Instrumentation for Detection of Exoplanets
  IV, ed. S.~B. {Shaklan}, 74400W, \dodoi{10.1117/12.826448}

\bibitem[{H. {K{\"u}hnle} {et~al.}(2025){K{\"u}hnle}, {Patapis},
  {Molli{\`e}re}, {Tremblin}, {Matthews}, {Glauser}, {Whiteford}, {Vasist},
  {Absil}, {Barrado}, {Min}, {Lagage}, {Waters}, {Guedel}, {Henning},
  {Vandenbussche}, {Baudoz}, {Decin}, {Pye}, {Royer}, {van Dishoeck},
  {{\"O}stlin}, {Ray}, \& {Wright}}]{Kuhnle2025}
{K{\"u}hnle}, H., {Patapis}, P., {Molli{\`e}re}, P., {et~al.} 2025,
  \bibinfo{title}{{Water depletion and $^{15}$NH$_{3}$ in the atmosphere of the
  coldest brown dwarf observed with JWST/MIRI},} \aap, 695, A224,
  \dodoi{10.1051/0004-6361/202452547}

\bibitem[{M. {K{\"u}rster} {et~al.}(2003){K{\"u}rster}, {Endl}, {Rouesnel},
  {Els}, {Kaufer}, {Brillant}, {Hatzes}, {Saar}, \& {Cochran}}]{Kuerster2003}
{K{\"u}rster}, M., {Endl}, M., {Rouesnel}, F., {et~al.} 2003,
  \bibinfo{title}{{The low-level radial velocity variability in Barnard's star
  (= GJ 699). Secular acceleration, indications for convective redshift, and
  planet mass limits},} \aap, 403, 1077, \dodoi{10.1051/0004-6361:20030396}

\bibitem[{M. {Kuzuhara} {et~al.}(2013){Kuzuhara}, {Tamura}, {Kudo}, {Janson},
  {Kandori}, {Brandt}, {Thalmann}, {Spiegel}, {Biller}, {Carson}, {Hori},
  {Suzuki}, {Burrows}, {Henning}, {Turner}, {McElwain}, {Moro-Mart{\'\i}n},
  {Suenaga}, {Takahashi}, {Kwon}, {Lucas}, {Abe}, {Brandner}, {Egner}, {Feldt},
  {Fujiwara}, {Goto}, {Grady}, {Guyon}, {Hashimoto}, {Hayano}, {Hayashi},
  {Hayashi}, {Hodapp}, {Ishii}, {Iye}, {Knapp}, {Matsuo}, {Mayama}, {Miyama},
  {Morino}, {Nishikawa}, {Nishimura}, {Kotani}, {Kusakabe}, {Pyo}, {Serabyn},
  {Suto}, {Takami}, {Takato}, {Terada}, {Tomono}, {Watanabe}, {Wisniewski},
  {Yamada}, {Takami}, \& {Usuda}}]{Kuzuhara2013}
{Kuzuhara}, M., {Tamura}, M., {Kudo}, T., {et~al.} 2013,
  \bibinfo{title}{{Direct Imaging of a Cold Jovian Exoplanet in Orbit around
  the Sun-like Star GJ 504},} \apj, 774, 11, \dodoi{10.1088/0004-637X/774/1/11}

\bibitem[{A.~M. {Lagrange} {et~al.}(2025){Lagrange}, {Wilkinson}, {M{\^a}lin},
  {Boccaletti}, {Perrot}, {Matr{\`a}}, {Combes}, {Beust}, {Rouan}, {Chomez},
  {Milli}, {Charnay}, {Mazevet}, {Flasseur}, {Olofsson}, {Bayo}, {Kral},
  {Carter}, {Crotts}, {Delorme}, {Chauvin}, {Thebault}, {Rubini}, {Kiefer},
  {Radcliffe}, {Mazoyer}, {Bodrito}, {Stasevic}, \& {Langlois}}]{Lagrange2025}
{Lagrange}, A.~M., {Wilkinson}, C., {M{\^a}lin}, M., {et~al.} 2025,
  \bibinfo{title}{{Evidence for a sub-Jovian planet in the young TWA 7 disk},}
  \nat, 642, 905, \dodoi{10.1038/s41586-025-09150-4}

\bibitem[{C.-P. {Lajoie} {et~al.}(2016){Lajoie}, {Soummer}, {Pueyo}, {Hines},
  {Nelan}, {Perrin}, {Clampin}, \& {Isaacs}}]{Lajoie2016}
{Lajoie}, C.-P., {Soummer}, R., {Pueyo}, L., {et~al.} 2016,
  \bibinfo{title}{{Small-grid dithers for the JWST coronagraphs},} in Society
  of Photo-Optical Instrumentation Engineers (SPIE) Conference Series, Vol.
  9904, Space Telescopes and Instrumentation 2016: Optical, Infrared, and
  Millimeter Wave, ed. H.~A. {MacEwen}, G.~G. {Fazio}, M.~{Lystrup},
  N.~{Batalha}, N.~{Siegler}, \& E.~C. {Tong}, 99045K,
  \dodoi{10.1117/12.2233032}

\bibitem[{B.~W.~P. {Lew} {et~al.}(2024){Lew}, {Roellig}, {Batalha}, {Line},
  {Greene}, {Murkherjee}, {Freedman}, {Meyer}, {Beichman}, {Alves de Oliveira},
  {De Furio}, {Johnstone}, {Greenbaum}, {Marley}, {Fortney}, {Young},
  {Leisenring}, {Boyer}, {Hodapp}, {Misselt}, {Stansberry}, \&
  {Rieke}}]{Lew2024}
{Lew}, B. W.~P., {Roellig}, T., {Batalha}, N.~E., {et~al.} 2024,
  \bibinfo{title}{{High-precision Atmospheric Characterization of a Y Dwarf
  with JWST NIRSpec G395H Spectroscopy: Isotopologue, C/O Ratio, Metallicity,
  and the Abundances of Six Molecular Species},} \aj, 167, 237,
  \dodoi{10.3847/1538-3881/ad3425}

\bibitem[{K.~L. {Luhman}(2014){Luhman}}]{Luhman2014}
{Luhman}, K.~L. 2014, \bibinfo{title}{{Discovery of a
  \raisebox{-0.5ex}\textasciitilde250 K Brown Dwarf at 2 pc from the Sun},}
  \apjl, 786, L18, \dodoi{10.1088/2041-8205/786/2/L18}

\bibitem[{K.~L. {Luhman} {et~al.}(2024){Luhman}, {Tremblin}, {Alves de
  Oliveira}, {Birkmann}, {Baraffe}, {Chabrier}, {Manjavacas}, {Parker}, \&
  {Valenti}}]{Luhman2024}
{Luhman}, K.~L., {Tremblin}, P., {Alves de Oliveira}, C., {et~al.} 2024,
  \bibinfo{title}{{JWST/NIRSpec Observations of the Coldest Known Brown
  Dwarf},} \aj, 167, 5, \dodoi{10.3847/1538-3881/ad0b72}

\bibitem[{M. {M{\^a}lin} {et~al.}(2024){M{\^a}lin}, {Boccaletti}, {Perrot},
  {Baudoz}, {Rouan}, {Lagage}, {Waters}, {G{\"u}del}, {Henning},
  {Vandenbussche}, {Absil}, {Barrado}, {Cossou}, {Decin}, {Glauser}, {Pye},
  {Olofsson}, {Glasse}, {Lahuis}, {Patapis}, {Royer}, {Scheithauer},
  {Whiteford}, {Serabyn}, {Choquet}, {Colina}, {Ostlin}, {Ray}, \&
  {Wright}}]{Malin2024}
{M{\^a}lin}, M., {Boccaletti}, A., {Perrot}, C., {et~al.} 2024,
  \bibinfo{title}{{Unveiling the HD 95086 system at mid-infrared wavelengths
  with JWST/MIRI},} \aap, 690, A316, \dodoi{10.1051/0004-6361/202450470}

\bibitem[{M. {M{\^a}lin} {et~al.}(2025){M{\^a}lin}, {Boccaletti}, {Perrot},
  {Baudoz}, {Rouan}, {Lagage}, {Waters}, {G{\"u}del}, {Henning},
  {Vandenbussche}, {Absil}, {Barrado}, {Charnay}, {Choquet}, {Cossou},
  {Danielski}, {Decin}, {Glauser}, {Pye}, {Olofsson}, {Glasse}, {Patapis},
  {Royer}, {Scheithauer}, {Serabyn}, {Tremblin}, {Whiteford}, {van Dishoeck},
  {Ostlin}, {Ray}, \& {Wright}}]{Malin2025}
{M{\^a}lin}, M., {Boccaletti}, A., {Perrot}, C., {et~al.} 2025,
  \bibinfo{title}{{First unambiguous detection of ammonia in the atmosphere of
  a planetary mass companion with JWST/MIRI coronagraphs},} \aap, 693, A315,
  \dodoi{10.1051/0004-6361/202452695}

\bibitem[{J. {Mang} {et~al.}(2026){Mang}, {Batalha}, {Morley}, {Wogan},
  {Mukherjee}, {Visscher}, {Marley}, {Fortney}, {Chubb}, {Gao}, \&
  {Malsky}}]{Mang2026}
{Mang}, J., {Batalha}, N.~E., {Morley}, C.~V., {et~al.} 2026,
  \bibinfo{title}{{PICASO 4.0: Clouds and Photochemistry in Climate Models of
  Brown Dwarfs and Exoplanets},} arXiv e-prints, arXiv:2602.22468,
  \dodoi{10.48550/arXiv.2602.22468}

\bibitem[{M.~S. {Marley} {et~al.}(2021){Marley}, {Saumon}, {Visscher}, {Lupu},
  {Freedman}, {Morley}, {Fortney}, {Seay}, {Smith}, {Teal}, \&
  {Wang}}]{Marley2021}
{Marley}, M.~S., {Saumon}, D., {Visscher}, C., {et~al.} 2021,
  \bibinfo{title}{{The Sonora Brown Dwarf Atmosphere and Evolution Models. I.
  Model Description and Application to Cloudless Atmospheres in Rainout
  Chemical Equilibrium},} \apj, 920, 85, \dodoi{10.3847/1538-4357/ac141d}

\bibitem[{C. {Marois} {et~al.}(2008){Marois}, {Macintosh}, {Barman},
  {Zuckerman}, {Song}, {Patience}, {Lafreni{\`e}re}, \& {Doyon}}]{Marois2008}
{Marois}, C., {Macintosh}, B., {Barman}, T., {et~al.} 2008,
  \bibinfo{title}{{Direct Imaging of Multiple Planets Orbiting the Star HR
  8799},} Science, 322, 1348, \dodoi{10.1126/science.1166585}

\bibitem[{C. {Marois} {et~al.}(2010){Marois}, {Zuckerman}, {Konopacky},
  {Macintosh}, \& {Barman}}]{Marois2010}
{Marois}, C., {Zuckerman}, B., {Konopacky}, Q.~M., {Macintosh}, B., \&
  {Barman}, T. 2010, \bibinfo{title}{{Images of a fourth planet orbiting HR
  8799},} \nat, 468, 1080, \dodoi{10.1038/nature09684}

\bibitem[{J.~V. {Martonchik} {et~al.}(1984){Martonchik}, {Orton}, \&
  {Appleby}}]{Martonchik1984}
{Martonchik}, J.~V., {Orton}, G.~S., \& {Appleby}, J.~F. 1984,
  \bibinfo{title}{{Optical properties of NH$_{3}$ ice from the far infrared to
  the near ultraviolet.},} \ao, 23, 541, \dodoi{10.1364/AO.23.000541}

\bibitem[{E.~C. {Matthews} {et~al.}(2024){Matthews}, {Carter}, {Pathak},
  {Morley}, {Phillips}, {P.~M.}, {Feng}, {Bonse}, {Boogaard}, {Burt},
  {Crossfield}, {Douglas}, {Henning}, {Hom}, {Ko}, {Kasper}, {Lagrange}, {Petit
  dit de la Roche}, \& {Philipot}}]{Matthews2024}
{Matthews}, E.~C., {Carter}, A.~L., {Pathak}, P., {et~al.} 2024,
  \bibinfo{title}{{A temperate super-Jupiter imaged with JWST in the
  mid-infrared},} \nat, 633, 789, \dodoi{10.1038/s41586-024-07837-8}

\bibitem[{E.~C. {Matthews} {et~al.}(2025){Matthews}, {Molli{\`e}re},
  {K{\"u}hnle}, {Patapis}, {Whiteford}, {Samland}, {Lagage}, {Waters}, {Tsai},
  {Zahnle}, {Guedel}, {Henning}, {Vandenbussche}, {Absil}, {Argyriou},
  {Barrado}, {Coulais}, {Glauser}, {Olofsson}, {Pye}, {Rouan}, {Royer}, {van
  Dishoeck}, {Ray}, \& {{\"O}stlin}}]{Matthews2025}
{Matthews}, E.~C., {Molli{\`e}re}, P., {K{\"u}hnle}, H., {et~al.} 2025,
  \bibinfo{title}{{HCN and C$_{2}$H$_{2}$ in the Atmosphere of a T8.5+T9 Brown
  Dwarf Binary},} \apjl, 981, L31, \dodoi{10.3847/2041-8213/adb4ec}

\bibitem[{W. McKinney(2010)McKinney}]{pandas_i}
McKinney, W. 2010, \bibinfo{title}{{D}ata {S}tructures for {S}tatistical
  {C}omputing in {P}ython,} in {P}roceedings of the 9th {P}ython in {S}cience
  {C}onference, ed. {S}t\'efan van~der {W}alt \& {J}arrod {M}illman, 56 -- 61,
  \dodoi{10.25080/Majora-92bf1922-00a}

\bibitem[{B.~E. {Miles} {et~al.}(2023){Miles}, {Biller}, {Patapis}, {Worthen},
  {Rickman}, {Hoch}, {Skemer}, {Perrin}, {Whiteford}, {Chen}, {Sargent},
  {Mukherjee}, {Morley}, {Moran}, {Bonnefoy}, {Petrus}, {Carter}, {Choquet},
  {Hinkley}, {Ward-Duong}, {Leisenring}, {Millar-Blanchaer}, {Pueyo}, {Ray},
  {Sallum}, {Stapelfeldt}, {Stone}, {Wang}, {Absil}, {Balmer}, {Boccaletti},
  {Bonavita}, {Booth}, {Bowler}, {Chauvin}, {Christiaens}, {Currie},
  {Danielski}, {Fortney}, {Girard}, {Grady}, {Greenbaum}, {Henning}, {Hines},
  {Janson}, {Kalas}, {Kammerer}, {Kennedy}, {Kenworthy}, {Kervella}, {Lagage},
  {Lew}, {Liu}, {Macintosh}, {Marino}, {Marley}, {Marois}, {Matthews},
  {Matthews}, {Mawet}, {McElwain}, {Metchev}, {Meyer}, {Molliere}, {Pantin},
  {Quirrenbach}, {Rebollido}, {Ren}, {Schneider}, {Vasist}, {Wyatt}, {Zhou},
  {Briesemeister}, {Bryan}, {Calissendorff}, {Cantalloube}, {Cugno}, {De
  Furio}, {Dupuy}, {Factor}, {Faherty}, {Fitzgerald}, {Franson}, {Gonzales},
  {Hood}, {Howe}, {Kraus}, {Kuzuhara}, {Lagrange}, {Lawson}, {Lazzoni}, {Liu},
  {Llop-Sayson}, {Lloyd}, {Martinez}, {Mazoyer}, {Quanz}, {Redai}, {Samland},
  {Schlieder}, {Tamura}, {Tan}, {Uyama}, {Vigan}, {Vos}, {Wagner}, {Wolff},
  {Ygouf}, {Zhang}, {Zhang}, \& {Zhang}}]{Miles2023}
{Miles}, B.~E., {Biller}, B.~A., {Patapis}, P., {et~al.} 2023,
  \bibinfo{title}{{The JWST Early-release Science Program for Direct
  Observations of Exoplanetary Systems II: A 1 to 20 {\ensuremath{\mu}}m
  Spectrum of the Planetary-mass Companion VHS 1256-1257 b},} \apjl, 946, L6,
  \dodoi{10.3847/2041-8213/acb04a}

\bibitem[{P. {Molli{\`e}re} {et~al.}(2022){Molli{\`e}re}, {Molyarova},
  {Bitsch}, {Henning}, {Schneider}, {Kreidberg}, {Eistrup}, {Burn}, {Nasedkin},
  {Semenov}, {Mordasini}, {Schlecker}, {Schwarz}, {Lacour}, {Nowak}, \&
  {Schulik}}]{Molliere2022}
{Molli{\`e}re}, P., {Molyarova}, T., {Bitsch}, B., {et~al.} 2022,
  \bibinfo{title}{{Interpreting the Atmospheric Composition of Exoplanets:
  Sensitivity to Planet Formation Assumptions},} \apj, 934, 74,
  \dodoi{10.3847/1538-4357/ac6a56}

\bibitem[{P. {Molli{\`e}re} {et~al.}(2025){Molli{\`e}re}, {K{\"u}hnle},
  {Matthews}, {Henning}, {Min}, {Patapis}, {Lagage}, {Waters}, {G{\"u}del},
  {J{\"a}ger}, {Zhang}, {Decin}, {Biller}, {Absil}, {Argyriou}, {Barrado},
  {Cossou}, {Glasse}, {Olofsson}, {Pye}, {Rouan}, {Samland}, {Scheithauer},
  {Tremblin}, {Whiteford}, {van Dishoeck}, {{\"O}stlin}, \&
  {Ray}}]{Molliere2025}
{Molli{\`e}re}, P., {K{\"u}hnle}, H., {Matthews}, E.~C., {et~al.} 2025,
  \bibinfo{title}{{Evidence for SiO cloud nucleation in the rogue planet PSO
  J318},} arXiv e-prints, arXiv:2507.18691, \dodoi{10.48550/arXiv.2507.18691}

\bibitem[{S.~E. {Moran} {et~al.}(2025){Moran}, {Lodge}, {Batalha}, {Ohno},
  {Vahidinia}, {Marley}, {Wakeford}, \& {Leinhardt}}]{virga_paper2}
{Moran}, S.~E., {Lodge}, M.~G., {Batalha}, N.~E., {et~al.} 2025,
  \bibinfo{title}{{Fractal Aggregate Aerosols in the Virga Cloud Code. I. Model
  Description and Application to a Benchmark Cloudy Exoplanet},} \apj, 994,
  116, \dodoi{10.3847/1538-4357/ae0583}

\bibitem[{S. {Mukherjee} {et~al.}(2023){Mukherjee}, {Batalha}, {Fortney}, \&
  {Marley}}]{Mukherjee2023}
{Mukherjee}, S., {Batalha}, N.~E., {Fortney}, J.~J., \& {Marley}, M.~S. 2023,
  \bibinfo{title}{{PICASO 3.0: A One-dimensional Climate Model for Giant
  Planets and Brown Dwarfs},} \apj, 942, 71, \dodoi{10.3847/1538-4357/ac9f48}

\bibitem[{S. {Mukherjee} {et~al.}(2024){Mukherjee}, {Fortney}, {Morley},
  {Batalha}, {Marley}, {Karalidi}, {Visscher}, {Lupu}, {Freedman}, \&
  {Gharib-Nezhad}}]{Mukherjee2024}
{Mukherjee}, S., {Fortney}, J.~J., {Morley}, C.~V., {et~al.} 2024,
  \bibinfo{title}{{The Sonora Substellar Atmosphere Models. IV. Elf Owl:
  Atmospheric Mixing and Chemical Disequilibrium with Varying Metallicity and
  C/O Ratios},} \apj, 963, 73, \dodoi{10.3847/1538-4357/ad18c2}

\bibitem[{K.~I. {{\"O}berg} {et~al.}(2011){{\"O}berg}, {Murray-Clay}, \&
  {Bergin}}]{Oberg2011}
{{\"O}berg}, K.~I., {Murray-Clay}, R., \& {Bergin}, E.~A. 2011,
  \bibinfo{title}{{The Effects of Snowlines on C/O in Planetary Atmospheres},}
  \apjl, 743, L16, \dodoi{10.1088/2041-8205/743/1/L16}

\bibitem[{P. {Pathak} {et~al.}(2021){Pathak}, {Petit dit de la Roche},
  {Kasper}, {Sterzik}, {Absil}, {Boehle}, {Feng}, {Ivanov}, {Janson}, {Jones},
  {Kaufer}, {K{\"a}ufl}, {Maire}, {Meyer}, {Pantin}, {Siebenmorgen}, {van den
  Ancker}, \& {Viswanath}}]{Pathak2021}
{Pathak}, P., {Petit dit de la Roche}, D.~J.~M., {Kasper}, M., {et~al.} 2021,
  \bibinfo{title}{{High-contrast imaging at ten microns: A search for
  exoplanets around Eps Indi A, Eps Eri, Tau Ceti, Sirius A, and Sirius B},}
  \aap, 652, A121, \dodoi{10.1051/0004-6361/202140529}

\bibitem[{A.~B.~T. {Penzlin} {et~al.}(2024){Penzlin}, {Booth}, {Kirk}, {Owen},
  {Ahrer}, {Christie}, {Claringbold}, {Esparza-Borges}, {L{\'o}pez-Morales},
  {Mayne}, {McCormack}, {Meech}, {Panwar}, {Powell}, {Sergeev}, {Taylor},
  {Wheatley}, \& {Zamyatina}}]{Penzlin2024}
{Penzlin}, A. B.~T., {Booth}, R.~A., {Kirk}, J., {et~al.} 2024,
  \bibinfo{title}{{BOWIE-ALIGN: how formation and migration histories of giant
  planets impact atmospheric compositions},} \mnras, 535, 171,
  \dodoi{10.1093/mnras/stae2362}

\bibitem[{M.~D. {Perrin} {et~al.}(2018){Perrin}, {Pueyo}, {Van Gorkom},
  {Brooks}, {Rajan}, {Girard}, \& {Lajoie}}]{Perrin2018}
{Perrin}, M.~D., {Pueyo}, L., {Van Gorkom}, K., {et~al.} 2018,
  \bibinfo{title}{{Updated optical modeling of JWST coronagraph performance
  contrast, stability, and strategies},} in Society of Photo-Optical
  Instrumentation Engineers (SPIE) Conference Series, Vol. 10698, Space
  Telescopes and Instrumentation 2018: Optical, Infrared, and Millimeter Wave,
  ed. M.~{Lystrup}, H.~A. {MacEwen}, G.~G. {Fazio}, N.~{Batalha}, N.~{Siegler},
  \& E.~C. {Tong}, 1069809, \dodoi{10.1117/12.2313552}

\bibitem[{M.~D. {Perrin} {et~al.}(2014){Perrin}, {Sivaramakrishnan}, {Lajoie},
  {Elliott}, {Pueyo}, {Ravindranath}, \& {Albert}}]{Perrin2014}
{Perrin}, M.~D., {Sivaramakrishnan}, A., {Lajoie}, C.-P., {et~al.} 2014,
  \bibinfo{title}{{Updated point spread function simulations for JWST with
  WebbPSF},} in Society of Photo-Optical Instrumentation Engineers (SPIE)
  Conference Series, Vol. 9143, Space Telescopes and Instrumentation 2014:
  Optical, Infrared, and Millimeter Wave, ed. J.~M. {Oschmann}, Jr.,
  M.~{Clampin}, G.~G. {Fazio}, \& H.~A. {MacEwen}, 91433X,
  \dodoi{10.1117/12.2056689}

\bibitem[{M.~W. {Phillips} {et~al.}(2020){Phillips}, {Tremblin}, {Baraffe},
  {Chabrier}, {Allard}, {Spiegelman}, {Goyal}, {Drummond}, \&
  {H{\'e}brard}}]{Phillips2020}
{Phillips}, M.~W., {Tremblin}, P., {Baraffe}, I., {et~al.} 2020,
  \bibinfo{title}{{A new set of atmosphere and evolution models for cool T-Y
  brown dwarfs and giant exoplanets},} \aap, 637, A38,
  \dodoi{10.1051/0004-6361/201937381}

\bibitem[{J. {Rameau} {et~al.}(2013){Rameau}, {Chauvin}, {Lagrange},
  {Boccaletti}, {Quanz}, {Bonnefoy}, {Girard}, {Delorme}, {Desidera}, {Klahr},
  {Mordasini}, {Dumas}, \& {Bonavita}}]{Rameau2013}
{Rameau}, J., {Chauvin}, G., {Lagrange}, A.-M., {et~al.} 2013,
  \bibinfo{title}{{Discovery of a Probable 4-5 Jupiter-mass Exoplanet to HD
  95086 by Direct Imaging},} \apjl, 772, L15,
  \dodoi{10.1088/2041-8205/772/2/L15}

\bibitem[{T.~L. {Roellig} {et~al.}(2004){Roellig}, {Van Cleve}, {Sloan},
  {Wilson}, {Saumon}, {Leggett}, {Marley}, {Cushing}, {Kirkpatrick}, {Mainzer},
  \& {Houck}}]{Roellig2004}
{Roellig}, T.~L., {Van Cleve}, J.~E., {Sloan}, G.~C., {et~al.} 2004,
  \bibinfo{title}{{Spitzer Infrared Spectrograph (IRS) Observations of M, L,
  and T Dwarfs},} \apjs, 154, 418, \dodoi{10.1086/421978}

\bibitem[{D. {Rouan} {et~al.}(2000){Rouan}, {Riaud}, {Boccaletti},
  {Cl{\'e}net}, \& {Labeyrie}}]{Rouan2000}
{Rouan}, D., {Riaud}, P., {Boccaletti}, A., {Cl{\'e}net}, Y., \& {Labeyrie}, A.
  2000, \bibinfo{title}{{The Four-Quadrant Phase-Mask Coronagraph. I.
  Principle},} \pasp, 112, 1479, \dodoi{10.1086/317707}

\bibitem[{N.~C. {Santos} {et~al.}(2004){Santos}, {Israelian}, \&
  {Mayor}}]{Santos2004}
{Santos}, N.~C., {Israelian}, G., \& {Mayor}, M. 2004,
  \bibinfo{title}{{Spectroscopic [Fe/H] for 98 extra-solar planet-host stars.
  Exploring the probability of planet formation},} \aap, 415, 1153,
  \dodoi{10.1051/0004-6361:20034469}

\bibitem[{B.~A. {Smith} \& R.~J. {Terrile}(1984){Smith} \&
  {Terrile}}]{Smith1984}
{Smith}, B.~A., \& {Terrile}, R.~J. 1984, \bibinfo{title}{{A Circumstellar Disk
  around {\ensuremath{\beta}} Pictoris},} Science, 226, 1421,
  \dodoi{10.1126/science.226.4681.1421}

\bibitem[{R. {Soummer} {et~al.}(2013){Soummer}, {Hines}, \&
  {Perrin}}]{Soummer2013}
{Soummer}, R., {Hines}, D., \& {Perrin}, M. 2013, {Simulations of target
  acquisition with MIRI four-quadrant phase mask coronagraph (I)},, Technical
  Report JWST-STScI-003063

\bibitem[{R. {Soummer} {et~al.}(2012){Soummer}, {Pueyo}, \&
  {Larkin}}]{Soummer2012}
{Soummer}, R., {Pueyo}, L., \& {Larkin}, J. 2012, \bibinfo{title}{{Detection
  and Characterization of Exoplanets and Disks Using Projections on
  Karhunen-Lo{\`e}ve Eigenimages},} \apjl, 755, L28,
  \dodoi{10.1088/2041-8205/755/2/L28}

\bibitem[{G. {Su{\'a}rez} \& S. {Metchev}(2022){Su{\'a}rez} \&
  {Metchev}}]{Suarez2022}
{Su{\'a}rez}, G., \& {Metchev}, S. 2022, \bibinfo{title}{{Ultracool dwarfs
  observed with the Spitzer infrared spectrograph - II. Emergence and
  sedimentation of silicate clouds in L dwarfs, and analysis of the full M5-T9
  field dwarf spectroscopic sample},} \mnras, 513, 5701,
  \dodoi{10.1093/mnras/stac1205}

\bibitem[{D. {Sudarsky} {et~al.}(2003){Sudarsky}, {Burrows}, \&
  {Hubeny}}]{Sudarsky2003}
{Sudarsky}, D., {Burrows}, A., \& {Hubeny}, I. 2003,
  \bibinfo{title}{{Theoretical Spectra and Atmospheres of Extrasolar Giant
  Planets},} \apj, 588, 1121, \dodoi{10.1086/374331}

\bibitem[{D. {Sudarsky} {et~al.}(2000){Sudarsky}, {Burrows}, \&
  {Pinto}}]{Sudarsky2000}
{Sudarsky}, D., {Burrows}, A., \& {Pinto}, P. 2000, \bibinfo{title}{{Albedo and
  Reflection Spectra of Extrasolar Giant Planets},} \apj, 538, 885,
  \dodoi{10.1086/309160}

\bibitem[{ {The pandas development team}(2021){The pandas development
  team}}]{pandas_ii}
{The pandas development team}. 2021, pandas-dev/pandas: Pandas 1.3.5, v1.3.5
  Zenodo, \dodoi{10.5281/zenodo.5774815}

\bibitem[{D.~P. {Thorngren} {et~al.}(2016){Thorngren}, {Fortney},
  {Murray-Clay}, \& {Lopez}}]{Thorngren2016}
{Thorngren}, D.~P., {Fortney}, J.~J., {Murray-Clay}, R.~A., \& {Lopez}, E.~D.
  2016, \bibinfo{title}{{The Mass-Metallicity Relation for Giant Planets},}
  \apj, 831, 64, \dodoi{10.3847/0004-637X/831/1/64}

\bibitem[{T. {Trifonov} {et~al.}(2020){Trifonov}, {Tal-Or}, {Zechmeister},
  {Kaminski}, {Zucker}, \& {Mazeh}}]{Trifonov2020}
{Trifonov}, T., {Tal-Or}, L., {Zechmeister}, M., {et~al.} 2020,
  \bibinfo{title}{{Public HARPS radial velocity database corrected for
  systematic errors},} \aap, 636, A74, \dodoi{10.1051/0004-6361/201936686}

\bibitem[{D. {Turrini} {et~al.}(2021){Turrini}, {Schisano}, {Fonte},
  {Molinari}, {Politi}, {Fedele}, {Pani{\'c}}, {Kama}, {Changeat}, \&
  {Tinetti}}]{Turrini2021}
{Turrini}, D., {Schisano}, E., {Fonte}, S., {et~al.} 2021,
  \bibinfo{title}{{Tracing the Formation History of Giant Planets in
  Protoplanetary Disks with Carbon, Oxygen, Nitrogen, and Sulfur},} \apj, 909,
  40, \dodoi{10.3847/1538-4357/abd6e5}

\bibitem[{M. {Vasist} {et~al.}(2025){Vasist}, {Mollire}, {K{\"u}hnle}, {Absil},
  {Louppe}, {Waters}, {G{\"u}del}, {Henning}, {Barrado}, {Decin}, {Pye}, \&
  {Tremblin}}]{Vasist2025}
{Vasist}, M., {Mollire}, P., {K{\"u}hnle}, H., {et~al.} 2025,
  \bibinfo{title}{{Panchromatic characterization of the Y0 brown dwarf WISEP
  J173835.52+273258.9 using JWST/MIRI},} arXiv e-prints, arXiv:2507.12264,
  \dodoi{10.48550/arXiv.2507.12264}

\bibitem[{P. Virtanen {et~al.}(2020)Virtanen, Gommers, Oliphant, Haberland,
  Reddy, Cournapeau, Burovski, Peterson, Weckesser, Bright, {van der Walt},
  Brett, Wilson, Millman, Mayorov, Nelson, Jones, Kern, Larson, Carey, Polat,
  Feng, Moore, {VanderPlas}, Laxalde, Perktold, Cimrman, Henriksen, Quintero,
  Harris, Archibald, Ribeiro, Pedregosa, {van Mulbregt}, \& {SciPy 1.0
  Contributors}}]{scipy}
Virtanen, P., Gommers, R., Oliphant, T.~E., {et~al.} 2020,
  \bibinfo{title}{{{SciPy} 1.0: Fundamental Algorithms for Scientific Computing
  in Python},} Nature Methods, 17, 261, \dodoi{10.1038/s41592-019-0686-2}

\bibitem[{C. {Visscher} {et~al.}(2006){Visscher}, {Lodders}, \&
  {Fegley}}]{Visscher2006}
{Visscher}, C., {Lodders}, K., \& {Fegley}, Jr., B. 2006,
  \bibinfo{title}{{Atmospheric Chemistry in Giant Planets, Brown Dwarfs, and
  Low-Mass Dwarf Stars. II. Sulfur and Phosphorus},} \apj, 648, 1181,
  \dodoi{10.1086/506245}

\bibitem[{G. {Viswanath} {et~al.}(2021){Viswanath}, {Janson}, {Dahlqvist},
  {Petit dit de la Roche}, {Samland}, {Girard}, {Pathak}, {Kasper}, {Feng},
  {Meyer}, {Boehle}, {Quanz}, {Jones}, {Absil}, {Brandner}, {Maire},
  {Siebenmorgen}, {Sterzik}, \& {Pantin}}]{Viswanath2021}
{Viswanath}, G., {Janson}, M., {Dahlqvist}, C.-H., {et~al.} 2021,
  \bibinfo{title}{{Constraints on the nearby exoplanet eps Indi Ab from deep
  near- and mid-infrared imaging limits},} \aap, 651, A89,
  \dodoi{10.1051/0004-6361/202140730}

\bibitem[{J.~J. {Wang} {et~al.}(2015){Wang}, {Ruffio}, {De Rosa}, {Aguilar},
  {Wolff}, \& {Pueyo}}]{Wang2015}
{Wang}, J.~J., {Ruffio}, J.-B., {De Rosa}, R.~J., {et~al.} 2015, pyKLIP: PSF
  Subtraction for Exoplanets and Disks,, Astrophysics Source Code Library,
  record ascl:1506.001

\bibitem[{M.~L. Waskom(2021)Waskom}]{seaborn}
Waskom, M.~L. 2021, \bibinfo{title}{seaborn: statistical data visualization,}
  Journal of Open Source Software, 6, 3021, \dodoi{10.21105/joss.03021}

\bibitem[{L. {Welbanks} {et~al.}(2024){Welbanks}, {Bell}, {Beatty}, {Line},
  {Ohno}, {Fortney}, {Schlawin}, {Greene}, {Rauscher}, {McGill}, {Murphy},
  {Parmentier}, {Tang}, {Edelman}, {Mukherjee}, {Wiser}, {Lagage}, {Dyrek}, \&
  {Arnold}}]{Welbanks2024}
{Welbanks}, L., {Bell}, T.~J., {Beatty}, T.~G., {et~al.} 2024,
  \bibinfo{title}{{A high internal heat flux and large core in a warm Neptune
  exoplanet},} \nat, 630, 836, \dodoi{10.1038/s41586-024-07514-w}

\bibitem[{N. {Whiteford} {et~al.}(2023){Whiteford}, {Glasse}, {Chubb},
  {Kitzmann}, {Ray}, {Phillips}, {Biller}, {Palmer}, {Rice}, {Waldmann},
  {Changeat}, {Skaf}, {Wang}, {Edwards}, \& {Al-Refaie}}]{Whiteford2023}
{Whiteford}, N., {Glasse}, A., {Chubb}, K.~L., {et~al.} 2023,
  \bibinfo{title}{{Retrieval study of cool, directly imaged exoplanet 51 Eri
  b},} \mnras, 525, 1375, \dodoi{10.1093/mnras/stad670}

\bibitem[{N.~F. {Wogan} {et~al.}(2025){Wogan}, {Mang}, {Batalha}, {Zahnle},
  {Mukherjee}, {Visscher}, {Fortney}, {Marley}, \& {Morley}}]{Wogan2025}
{Wogan}, N.~F., {Mang}, J., {Batalha}, N.~E., {et~al.} 2025,
  \bibinfo{title}{{The Sonora Substellar Atmosphere Models. V. A Correction to
  the Disequilibrium Abundance of CO$_{2}$ for Sonora Elf Owl},} Research Notes
  of the American Astronomical Society, 9, 108,
  \dodoi{10.3847/2515-5172/add407}

\bibitem[{J. {Xuan} {et~al.}(2025){Xuan}, {Ruffio}, {Sanghi}, {Zhang},
  {Horstman}, {Knutson}, {Madurowicz}, {Marois}, {Mawet}, {Oppenheimer},
  {Thompson}, \& {Wang}}]{Xuan_jwstgo8714}
{Xuan}, J., {Ruffio}, J.-B., {Sanghi}, A., {et~al.} 2025, {Combining isotopic
  and elemental abundances to unveil the formation and accretion history of a
  cold Jupiter},, JWST Proposal. Cycle 4, ID. \#8714

\bibitem[{M. {Zechmeister} {et~al.}(2013){Zechmeister}, {K{\"u}rster}, {Endl},
  {Lo Curto}, {Hartman}, {Nilsson}, {Henning}, {Hatzes}, \&
  {Cochran}}]{Zechmeister2013}
{Zechmeister}, M., {K{\"u}rster}, M., {Endl}, M., {et~al.} 2013,
  \bibinfo{title}{{The planet search programme at the ESO CES and HARPS. IV.
  The search for Jupiter analogues around solar-like stars},} \aap, 552, A78,
  \dodoi{10.1051/0004-6361/201116551}

\end{thebibliography}
\bibliographystyle{aasjournalv7}

\end{document}